\documentclass[sigconf]{acmart}

\usepackage{graphicx}

\usepackage{lscape}
\usepackage{xcolor}
\usepackage{lmodern}
\usepackage{eurosym}
\usepackage{url}
\usepackage{tabularx}

\copyrightyear{2022} 
\acmYear{2022} 
\setcopyright{rightsretained} 
\acmConference[ICSE '22]{44th International Conference on Software Engineering}{May 21--29, 2022}{Pittsburgh, PA, USA}
\acmBooktitle{44th International Conference on Software Engineering (ICSE '22), May 21--29, 2022, Pittsburgh, PA, USA}
\acmDOI{10.1145/3510003.3510100}
\acmISBN{978-1-4503-9221-1/22/05}

\begin{document}

\title{What Makes Effective Leadership in \mbox{Agile Software Development Teams?}}

\author{Lucas Gren}
\affiliation{%
  \institution{Volvo Cars and Chalmers $|$ University of Gothenburg}
    \city{Gothenburg}
  \country{Sweden}}
  \affiliation{%
  \institution{Blekinge Institute of Technology}
  \city{Karlskrona}
  \country{Sweden}}
\email{lucas.gren@lucasgren.com}

\author{Paul Ralph}
\affiliation{%
  \institution{Dalhousie University}
  \city{Halifax}
  \country{Canada}}
\email{paulralph@dal.ca}

\begin{abstract}
  Effective leadership is one of the key drivers of business and project success, and one of the most active areas of management research.  But how does leadership work in agile software development, which emphasizes self-management and self-organization and marginalizes traditional leadership roles?  To find out, this study examines agile leadership from the perspective of thirteen professionals who identify as agile leaders, in different roles, at ten different software development companies of varying sizes. Data from semi-structured interviews reveals that leadership: (1) is dynamically shared among team members; (2) engenders a sense of belonging to the team; and (3) involves balancing competing organizational cultures (e.g. balancing the new agile culture with the old milestone-driven culture). In other words, agile leadership is a property of a team, not a role, and effectiveness depends on agile team members' identifying with the team, accepting responsibility, and being sensitive to cultural conflict.
\end{abstract}



\maketitle

\section{Introduction}\label{sec:introduction}
Leadership is one of the key drivers of organizational and project success~\cite{bass2006transformational}. However, agile software development involves a cultural change~\cite[cf.][]{iivari,tolfo2008} that flattens the organizational hierarchy, de-emphasizes managers and empowers teams~\cite{moe2009overcoming}. One might therefore wonder whether there is leadership in agile teams at all. 

However, management is different from leadership.  ``Leadership [involves] the articulation of an organizational vision, introduction of major organizational change, providing inspiration, and dealing with highly stressful and troublesome aspects of the external environments of organizations''~\cite[p. 444]{house1997social}. In contrast, ``Management consists of implementing the vision and strategy provided by leaders, coordinating and staffing the components of organizations, administering the infrastructures of organizations, and handling the day-to-day problems that inevitably emerge in the process of strategy and policy implementation and ongoing organizational functioning''~\cite[p. 445]{house1997social}. 

Put more simply, ``leadership is getting group members to achieve the group's goals''~\cite[p. 315]{hogg2014sp}. \textit{Leadership is a kind of work.} Even in self-organized and self-managed agile teams, someone is doing leadership. Indeed, many team members can dynamically share the leadership work of driving the team toward its goals. 
 
Research on leadership in software engineering is scarce. One notable exception is Kalliamvakou et al.'s study of Microsoft managers, which concluded that ``we still don't know what to look for in a great software engineering manager, and how to further develop their skills to support the teams they manage''~\cite{kalliamvakou2017makes}. A recent systematic review concluded that ``agile leadership research needs further attention and that more empirical studies
are needed to better understand agile leadership in general''~\cite{modi2020leadership}. This motivates the following research question.

\smallskip
{\narrower \noindent \textit{\textbf{Research question:} What is the nature of leadership in agile 
software development, from the perspective of professionals who identify as agile leaders?} \par}
\smallskip

\begin{table*}
\footnotesize
\renewcommand{\arraystretch}{1.3}
\caption{Participant Information}
\label{tab:participants}
\begin{tabular}{llp{23mm}rp{40mm}p{27mm}p{33mm}}
\bfseries PID & \bfseries Org. & \bfseries Industry & \bfseries Employees & \bfseries Role  & \bfseries Prior Method & \bfseries Reason for agile\\
\hline
A1  &A&  Consumer packaged goods & 26,000 & Agile coach  & Waterfall &Engagement, job satisfaction, and quality\\
\hline
B1 &B& Retail business & 16,000 & Project manager lead & Waterfall & Improve business value\\
\hline
C1 & C  &  Automatic identification\slash  data capture & 1,800 & Project portfolio management responsible & Ad hoc process & Project priorities not personal interest\\
\hline
D1 & D&  Industrial supply distribution & 18,400 & Project Manager (initiator of agile). & Waterfall & Innovation in value delivery\\
\hline
E1 & E & Personal care & 56,000  & Team leader sales and distr. & Waterfall & Improve the company\\
\hline
E2 & E & Personal care & 56,000  & Lead of ~25 project managers & Waterfall & Improve the company\\
\hline
F1 & F &Oil and gas & 82,000 & Project manager (project execution) & Waterfall & Read about agile  methods\\
\hline
G1 & G &  Software  & 74,400 & Scrum Master\slash Project Manager & Started agile &Speed\\
\hline
G2 & G &  Software  & 74,400 & Scrum Master of two teams& Started agile &Speed\\
\hline
G3 & G &  Software  & 74,400 & Scrum Master  & Started agile &Speed\\
\hline
H1 & H & Online media and social networking & 5,000  & Scrum Master\slash manager in first agile team  & A culture of guessing what users liked & Better and faster feedback\slash solve org. problem \\
\hline
I1 & I & Training and education & 100 & Multi-type supporting role & Similar to agile  &Started as agile\\
\hline
J1 & J &  Software consultancy & 35 & Founder & Agile principles with flat org. structure & Started as agile\\
\hline
\end{tabular}
\end{table*}

Here, agile software development refers to both the ideology laid out in the agile manifesto~\cite{fowler2001} and the specific methodologies and practices that embody that ideology (e.g.\ Extreme Programming~\cite{beck1999embracing}, Scrum~\cite{schwaber}, Dual-Track Development~\cite{sedano2020dual}). 

To address this research question, we conducted a qualitative survey (a series of interviews) of thirteen self-described agile leaders from ten different organizations (see Section~\ref{sec:methodology}). Section~\ref{sec:results} presents our findings. Next, we describe existing research related to our findings (Section~\ref{sec:RelatedWork}), and then integrate our findings with this work (Section~\ref{sec:discussion}). Section~\ref{vts} describes the study's limitations and Section~\ref{con} concludes the paper with some thoughts on future work.

\section{Method}\label{sec:methodology}
This section explains our approaches to data collection and analysis. 
Since our research question concerns the perspectives of agile leaders, we employ a qualitative survey~\cite{ralph2020acm} targeting professionals who identify as agile leaders.

\subsection{Participants}\label{participants}
The first author approached industry contacts---with whom he had no direct relationship or conflict of interest---to recruit participants who identify as agile leaders (e.g., Scrum Masters, founders, agile coaches, and various project managers). We focused on agile leaders rather than other professionals because we wanted to understand the leaders' perspective. Of course, comparing the leaders' perspective with software professionals who explicitly do not identify as leaders is a potential extension. We also wanted participants of different ages, from both small and large companies, and from different countries, to maximize sample diversity---a key aspect of representativeness~\cite{baltes2020sampling}. Moreover, we want to investigate components of agile leadership that are similar across these contextual characteristics. 

More formally, the recruitment criteria were: (1) Employees who identify as agile leaders, (2) Companies of different size, (3) Mix of age and gender, and (4) Mix of countries and national cultures.

Nine men and four women between 25- and 50-years-old participated. Six participants were located in the USA, three in Brazil, two in Germany, one in the UK, and one in Tunisia. All participants indicated that their teams use an agile approach. Participants roles, industries, organization sizes and methodological contexts varied (see Table~\ref{tab:participants}). SAP America Inc.\ mediated some of the contacts to the companies; four of the participants were working on Agile SAP implementations, meaning that they use agile methods in customizing an SAP product for an organization~\cite{sean}. All four also worked on other, non-SAP projects.

\subsection{Data Collection}
The first author completed 13, 45--90 minute, semi-structured interviews: ten via teleconference; three face-to-face. He began by introducing himself and the project, getting permission to record the interview and emphasizing how participants' identities would be protected. The complete interview guide is available in Supplementary Material (see \nameref{suppl}). All interviews were transcribed word-for-word. 


\subsection{Data Analysis}
The data was analyzed using ATLAS.ti\footnote{\url{https://atlasti.com/}} following the steps recommended by Braun et al.~\cite{braun2006using}: 
\begin{enumerate}
    \item read each transcript
    \item highlight all statements broadly related to concept under investigation (i.e.\ \emph{leadership} in this case)
    \item sort the highlighted statements into categories
    \item name each category
    \item for each category, re-read all of its statements together; reassess category cohesion and name
\end{enumerate}

The first author began by reading all the transcripts and tagging all quotations related to leadership in Atlas.ti---176 in all. Next, quotations were grouped into categories and the categories were iteratively refined. 

We applied the two concepts of saturation as described by Aldiabat and Navenec~\cite{aldiabat2018data}, i.e., \emph{code} and \emph{meaning} saturation. We recognized \emph{code} saturation when no new codes emerged from the last few interviews. After this step, the second author audited the categories for coherence. The categories were then revisited and rephrased again. We recognized \emph{meaning} saturation when we understood our themes well enough to see their relationship to existing theories in social psychology (described in Section~\ref{sec:RelatedWork}). 
We then integrated our categories back into existing theories (see Section \ref{sec:discussion}). The three themes described next emerged from this synthesis of our categories with existing research. Tables showing examples for all emerging themes and categories are included in Supplementary Material (see \nameref{suppl}). Two categories---``communication'' and ``staffing problems''---were dropped as out of scope for this paper. The reason they were out of scope was that the participants who mentioned communication and staffing problems discussed them as general issues, not directly related to effective agile leadership.

\section{Findings}\label{sec:results}
\begin{figure*}[ht]
\centering
\includegraphics[width=0.9\linewidth]{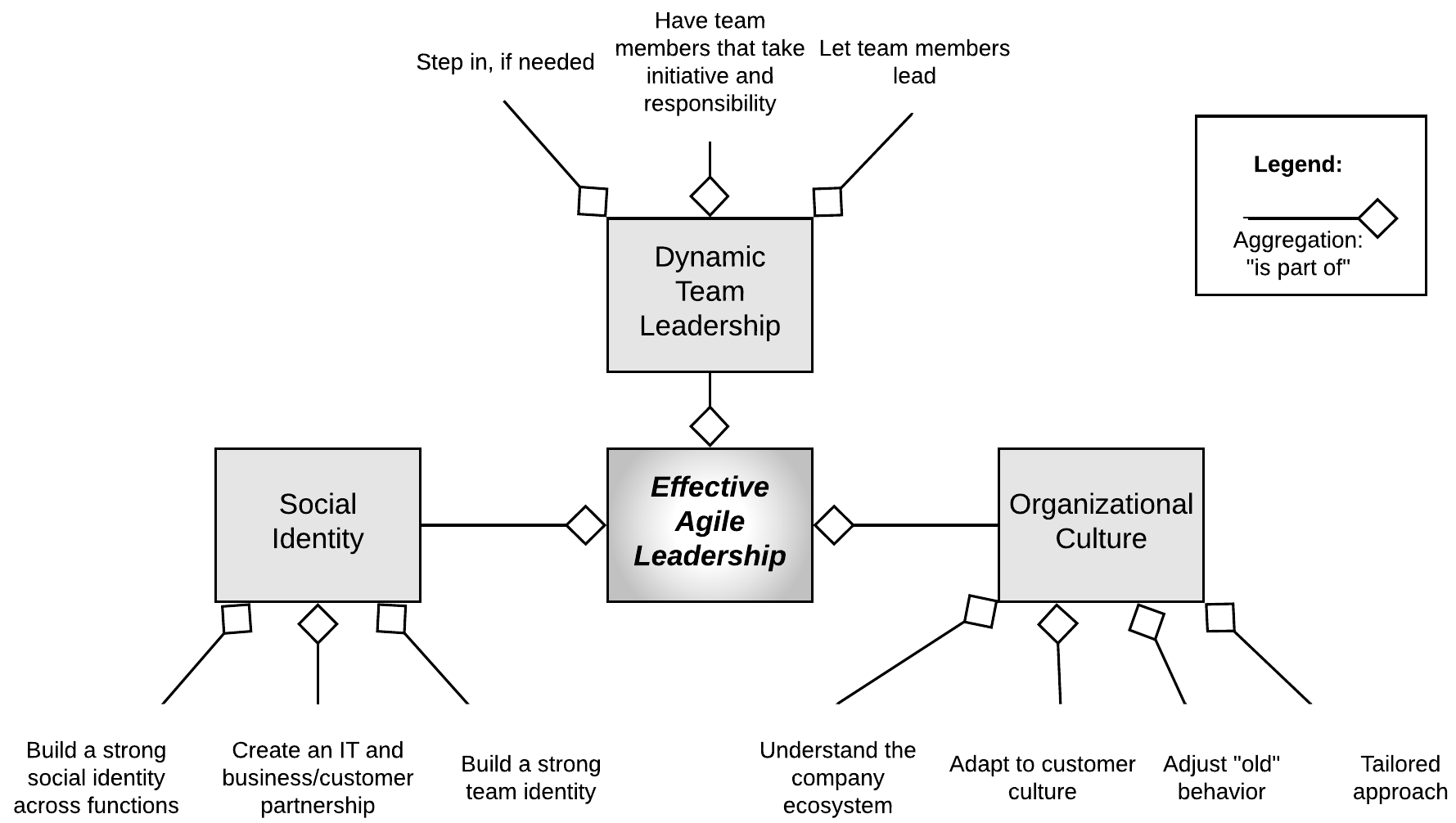}
\caption{Components of effective leadership in agile software development teams.}
\label{fig_sim}
\end{figure*}

Figure~\ref{fig_sim} and Table~\ref{tab:result}  summarize our findings. Briefly, agile leadership comprises three dimensions: \emph{dynamic team leadership}, \emph{social identity} and \emph{organizational culture}.

\begin{table*}
\footnotesize
\renewcommand{\arraystretch}{1.3}
\caption{Overview of Themes and Categories with Example Quotations}
\label{tab:result}
\begin{tabularx}{\linewidth}{p{17mm}p{25mm}X}
\bfseries Theme & \bfseries Category & \bfseries Examples of supporting quotations\\
\hline
Dynamic team \mbox{leadership}  & Have team members that take initiative and responsibility & ``Another guy said, `well you won't have the time, so I'm the Scrum Master now' and that was it \ldots they don't see the Scrum Master as a person, they see it as a role that they all can take.'' (Participant I1) 

``Teams that are doing it right \ldots are really thinking about being collaborative and self-organized and all those things. They become passionate about it and really want to do more of it.'' (Participant C1)\\
  & Let team members lead & ``I guess we were just really lucky with the team members we had. This was basically also my objective; careful! Don't become a project manager, don't start planning stuff, don't start assigning or defining tasks. Stay out of that!'' (Participant G3)
  
  ``There are teams that have a strong leader, and there the problem is another one which is this whole idea of having someone telling everybody what to do. For me, it's the leadership will make it the best or the worst.'' (Participant J)\\
  & Step in, if needed & ``Most of the time we have to interfere as leaders in order to help the team come together and make a better decision.'' (Participant J1)

  ``I could [force the teams to try a practice]. And I did. \ldots And if they don't like it, they're free to go back to the old one.'' (Participant I1)\\
  
\hline
Social identity  & Build a strong social identity across functions & ``A new resource from finance for example didn't used to have much to do with the other functions. Now with these meetings and planning, they are all one team working together and cheer each other's aspects.'' (Participant F1)

``It's very easy to see that, well, `my part in the project is just coding,' but that's not quite true. They [the team] get less cohesive by having a clear separation of roles.'' (Participant H)\\
    & Create an IT and business\slash customer partnership & ``The business partnership and engagement you create is absolutely tremendous compared to the traditional approach!'' (Participant B1)
    
    ``So they can again, work to be a part of and share the accountability [in agile]. I mean, for a whole variety of reasons that involvement of the business partner as a part of the team, not them versus us.'' (Participant D)\\
    & Build a strong team identity & ``It's interesting because when we deploy software in the middle of the day, the [people] here actually call the other units and say `ok, we're stopping the software for a couple of minutes for deploying.' So then everyone stops. We could do a night deploy, but we prefer to do that.'' (Participant H1)
    
    ``The Scrum part takes between seven to ten minutes. Then it takes another five to ten minutes to fill out the time, and it's in the spreadsheet, so then I got to the tab that shows the burn down. People wait for that! Like: `Oooh let's see what the burn-down is today!' And then the third part is just an open forum, an open meeting, and people sort of take care of things that will only take a short amount of time to discuss.'' (Participant G1)\\
    
\hline
Organizational culture   & Tailored approach & ``I have a kind of waterfall-like representation of our sprints and the main milestones and I show that to them. But then I just take care of all the requirements from the organization, you know, at a higher level. The team doesn't have to worry about that stuff.'' (Participant G1)

``What one individual pilot project starts to practice, you always have difficulty scaling that, because that's what they were doing and got really good at doing exactly that specific to that project. But that's not necessarily well translated into practice format across the enterprise. That needs to be more of the fundamentals and the process behind it.''(Participant A1)\\
    & Understanding the company ecosystem & ``If you haven't brought those people along in your agile journey, then subsequently you will have that hurdle to overcome because of the different culture and behaviors.'' (Participant D1)
    
    ``To some people it's a subtle difference and they sometimes have a hard time grasping why we don't just have one list for finance, one list for sales, ops, etc. But it has been a key drive for us since we have shared resources, to have one cross-functional list.'' (Participant F1)\\
    & Adapt to customer culture & ``Often, they don't want changes as often as we could deliver them.'' (Participant C1)
    
    ``We used to work mostly with start-up companies, and with start-up companies, that's a big discussion because, with a tight budget, why would you have 17 layers of tests?'' (Participant J1)\\
    &  Adjust ``old'' behavior & ``Allowing the team to become a high performance team without the problem of having a lot of politics involved in the execution of the project.'' (Participant E2)
    
    ``Waterfall approach provides some very familiar and comfortable handrails, called milestones, which everybody grasps and clings to for life.'' (Participant G1)\\
\hline
`\end{tabularx}
\end{table*}

\smallskip{}

\begin{center}
\noindent\framebox{%
  \begin{minipage}{0.95\linewidth}
    \textbf{In other words, we found that agile teams:}
    \begin{enumerate}
        \item \textbf{share leadership work rather than having an individual leader;}
        \item \textbf{build a sense of belonging and common purpose across and within organizational functions;}
        \item \textbf{adapt their processes to the different cultures which exist simultaneously within a single organization.}
    \end{enumerate}
  \end{minipage}
}    
\end{center}

\subsection{Dynamic Team Leadership}
Ten out of thirteen interviewees highlighted that agile team leadership is only effective when it is shared by the team. All the interviewees saw themselves as an active part of obtaining a shared-leadership dynamic.

\subsubsection{Have Team Members that Take Initiative and Responsibility}
Interviewees emphasized having a team where everybody could take initiative. 
One interviewee stated that high-performing individuals do not make good agile team members because teamwork skills are more important than individual performance. Interviewees also described how team members who were ``initiators'' could enroll others in leadership efforts and help make shared leadership a team norm. 

Another interviewee stressed how a single team member could ruin the team's effectiveness due to passiveness and not being able to create an active work-mode. This finding mirrors prior research on dysfunctional teams~\cite{felps2006and}. 

Interviewees felt that complaining was not necessarily detrimental to the team's effectiveness as long as the team members felt like they were in charge of making a change for the better. Complaining seems more infectious and destructive if teams do not have autonomy because they tend to complain about factors beyond their control. 

One interviewee explained that teams need some sort of leadership, whether or not they are agile, to solve problems effectively. This interviewee saw the positive effects of having all team members be leaders somehow but also stressed that many people in industry are used to having managers that assume responsibility for project failures. Getting people to think differently about leadership as shared was therefore challenging. Interviewees' perspectives on shared leadership are complicated by the inability to divide projects neatly into agile vs. plan-driven---some officially plan-driven projects were described as being quite agile and vice versa.


The interviewees also stressed that there was often a difference in that agile methods have more onerous expectations for all team members doing leadership work than plan-driven methods. One interviewee was skeptical that agile is good for all projects because agile methods assume that team members are willing to accept responsibility, take initiative, and make decisions.  

People with a mindset of doing leadership themselves appear to focus more on agile principles and less on implementing user stories as quickly as possible. Teams with such members focused more on being collaborative, self-organized, and seemed more passionate about their work. 

Another interviewee attributed the team members' active participation in driving the teamwork to the agile approach to projects. The motivation to lead was therefore not only an a priori property of individual team members but also an emerging team norm. The interviewee emphasized this distinction because it implies that it is possible to activate leadership behavior by setting other group norms~\cite{teh}. 

\subsubsection{Let Team Members Lead}
Having members who want to take initiative is necessary but insufficient. Eight interviewees indicated that more traditional forms of leadership (i.e.\ managers) surrounding a team could undermine self-leadership. Teams were described as needing to feel heard and autonomy not only for technical or engineering practices but also for team improvements and other organizational issues. 

Another key skill that surfaced was a leader's ability to take a step back when leadership can be shared. This was described as being tricky because guidance was often needed early on and it could be difficult to spot when an individual or a team is ready for increased responsibility. One interview explained that, especially when leadership is a formal role, the people in that role need to actively avoid taking over the leadership of the team. Some interviewees believe that strong, formal leadership roles prevent self-organization and shared leadership. 

It was sometimes unclear to interviewees whether individual leadership remained because it was needed, or for other reasons. One interviewee described a situation where their team's effectiveness was reduced by the manager's unwillingness to share leadership. Another stressed that leadership, when not shared, can become destructive. 

\subsubsection{Step In, If Needed}
Two of the interviewees explicitly stated that, even if they strove for self-organization, an experienced supervisor sometimes needed to intervene. One person described this as being valued by the teams because the ways in which the supervisor could intervene were clear. The other interviewee expressed frustration in having to help teams make better decisions. They felt that the team should be ready to accept responsibility faster, but were not, and therefore tried to help. Having the support of a senior employee was also described as helpful since team members could validate issues with that person before feeding the issues back to the team itself. 

\subsection{Social Identity}
Of the ten organizations where our interviewees work, organizations G, I, and J, embraced agile methods from their founding. In contrast, organizations A--F and H are large companies with a long history of milestone-driven software projects. They were organized around strong functional silos (e.g. marketing, production, software development, accounting \& finance). Many participants were therefore leading transitions from a functional organizational hierarchy to a project-based organization, or leading cross-functional agile teams within companies that were organized around organizational functions.\footnote{Throughout this paper, ``functions'' refers to organizational functions (e.g. marketing, accounting).}  

All of the participants viewed cross-functional teams with high levels of team spirit as advantageous because it helps more employees take responsibility. The participants explained that they obtained higher levels of team spirit, not only for their own work context and field of expertise but also for other important organizational functions (e.g.\ finance). None of the participants used the term ``social identity''; however, they all discussed consolidating, stabilizing, and anchoring different social categories (e.g.\ business people, developers, testers). A person's social identity is the collection of social categories with which that person identifies~\cite{hornsey2008social}. 

\subsubsection{Build a Strong Social Identity Across Functions}

One of the most highlighted benefits of agile software development is increased team spirit; that is a greater sense of team-belonging and social cohesion. One interviewee explained how, as members come together from different functional backgrounds, a strong sense of team spirit (i.e. feelings of belonging) helps them on-board, integrate with the team and begin contributing. As team membership becomes part of the new team member's social identity, they are more willing to accept responsibility and do leadership work. 

Another participant highlighted that a separation of roles within the team decreased team cohesion. They therefore focused on getting team members to realize the value of sharing all aspects of software delivery instead of focusing on their individual expertise. By sharing in different tasks, team members expanded their social identity to include belonging to a team that was defined by its \textit{goal}, rather than its \textit{roles}. 

\subsubsection{Create an IT and Business\slash Customer Partnership}
Moreso than previous approaches, agile software development focuses on social identities. Transitioning to agile necessitates defining one's work identity based on social categories other than roles. In other words, agilists are supposed to identify with their cross-functional team instead of, or at least in addition to, their role. For example, employees should say `I'm part of the \textit{Mass Effect} team,' or `I'm on project Urithiru' not `I'm a front-end developer' or `I'm a requirements analyst'. Interviewees consistently described how their teams obtained a common social identity by sitting and working together. 

Participants emphasized how leaders expanded their team members' responsibilities by communicating who was responsible for what. One participant described attempting to enroll customers into the team's shared identity. They felt that projects in which customers and developers felt like they were on the same team were more successful. 

Two participants described cases when they did not succeed in creating a common social identity with business people or other external customers. Such situations were described as leading to team members not feeling prioritized or sharing accountability for the work being done. 

\subsubsection{Build a Strong Team Identity}
Social identity building within the teams was also described as important. All practitioners highlighted the importance of the social ceremonies (often taken from agile ceremonies like a retrospective meeting or daily stand-ups~\cite{so}) in creating a shared social category that became a part of the individuals' social identity. One participant described a ceremony that was possible to conduct more efficiently, but the team preferred deploying code in the middle of the day so that everyone could get together and get an important sense of belonging in the team.

Building a strong and common social identity was also described as being easier for co-located teams, which is consistent with prior research~\cite{korhonen2013evaluating}. Two of the participants described putting extra effort into online ceremonies to engender a sense of belonging in their distributed teams.

\subsection{Organizational Culture}
Many large companies do not really want an agile transition. They just want to increase speed and productivity by adopting some elements of agile while retaining their functional silos and tyrannical power hierarchies. Agile leaders, not only within these large companies but also within smaller organizations beholden to them, therefore have to balance conflicting organizational cultures---the new agile culture and the old, milestone-driven culture. This balancing involves tailoring the agile approach, understanding the company ecosystem, adapting to customer culture, and adjusting ``old'' behaviors. All of the participants repeatedly contrasted effective leadership in an agile context to other leadership behavior they had also observed or experienced.

\subsubsection{Tailored Approach}
One common topic was the difficulty of scaling up small agile pilot projects to other contexts, even within the same company. The larger companies tended to want a standard set of agile practices, but participants reported that the diversity of contexts hinders a standardized approach. Some participants manage this conflict between diversity and standardization by focusing on the agile principles while tailoring the practices for each, or a few, teams.  

One strategy described by three participants was to translate the agile principles into the company's context rather than rigidly implementing agile practices and rituals. One participant described tweaking the waterfall model to make it more incremental as a first step toward agile. Another interviewee claimed that agile was not appropriate for some phases in their development---it simply did not fit where the company was at the time of the interview. Returning to the idea of balancing opposing cultures, another participant suggested that agile leaders should shield the team from the control needs of their organizations so that the team can focus on delivering value.  

\subsubsection{Understand the Company Ecosystem}
With large-scale agile, one participant highlighted the issues with having only one product owner for each team since that person struggled to set priorities for the teams since that person alone could not balance many different functions (e.g. stakeholders, finance, line management). They instead focused on the function and purpose of product ownership, and created a business council where up to seven people from different organizational functions set the priorities together. One representative from each department (e.g.\ one from finance, one from marketing, one from operations, etc.) met every three weeks to review all projects and their previous priorities. The council then created one single list of company priorities across the organizational functions and, through the creation of that list, removed silos and forced collaboration across the company. This appeared to improve consensus that IT was working on the most important projects for the company.

Another participant emphasized that some companies they worked with as customers did not understand that transitioning to agile development entails a cultural change. Balancing two conflicting cultures was described as very important within the participant's own organization, but also to customize teams for customer culture. This internal adaptation required much effort because customers that did not incorporate and focused on agile principles did not obtain as much added values as the ones that did, which was described as frustrating but necessary. 

Participants not only had to see to the teams' need to contextualize but also to balance other functions within the same company, all of them with a different culture. One example being that a development team could have an agile financial department integrated into their project so that they could make rapid changes to what was being built, while others had a more milestone driven external financial department hindering their agility. 

\subsubsection{Adapt to Customer Culture}
All participants discussed balancing their internal organizational cultures with their customers' organizational cultures. One participant claimed that their team could deliver updates much more often than the customer could handle them. Therefore, the participant felt that they had to be less agile to adapt to their customers. (The implication that agility is determined by release schedule is questionable~\cite{wufka2015explaining} but many participants seem to think agile means delivering fast and often, rather than being responsive to change). 

One participant said that agile teams and their stakeholders meet and synchronize their expectations instead of specifying requirements and then controlling delivery. They felt that this synchronization nurtured more discipline and commitment in stakeholders.  


Another cultural tension concerned contracting. One participant complained that traditional fixed-price contracts hampered agile ways of working, which is consistent with previous research~\cite[e.g.][]{jorgensen2017direct}. The contracts instead defined what type of collaboration was possible with the customers, and partly determined how the two different cultures could be balanced. What this participant described was contracting based on the agile principles, i.e.\ fixed time, fixed cost, but flexible scope. They also had fully dedicated teams for each project and avoided developers in multiple teams. 

\subsubsection{Adjust ``Old'' Behavior}
The dialectic tension between the old, milestone-driven culture and the new agile culture, including power, values, and principles of working, exists at both the organizations individual levels. All participants described old habits and old ways of working resurfacing continuously in the new agile environments. Agile leaders therefore felt that they needed to guard new work principles and continuously adjust deviating behaviors.  

One participant explained that they were so used to the old ways of working that many employees struggled to follow the reasoning behind the new ways of working. They described partially losing control of their project and having to let go of aspects of control that used to be key for them to know how to do their job.  

All participants described working hard to change behavior. One participant claimed that changing the behavior of people in their organization was the most challenging part of creating a more customer value driven company.

All interviewees mentioned how effective leadership meant letting go of control and reporting, while trusting teams and focusing on incremental delivery of customer value. Participants reported varying results with about half of their companies having reasonably successful translations while the other half were still struggling. 

\subsection{Summary}
In summary, participants identified three broad capabilities necessary for effective leadership in modern software development:
\begin{enumerate}
    \item The ability to share responsibility and leadership work with other team members.
    \item The ability to get team members from different functional silos to identify as members of a cross-functional team.
    \item The ability to manage ongoing tension between the competing logics of the old, milestone-driven culture and the new agile culture.
\end{enumerate}

\section{Related Work}\label{sec:RelatedWork}
This section summarizes existing literature on leadership in software engineering and social psychology. Then, in Section \ref{sec:discussion}, we will integrate the findings from Section \ref{sec:results} with the existing theory and research reviewed in this section. 

\subsection{Research on Leadership in Software Engineering}\label{selead}
In software development, \textit{agility} has at least three meanings:
\begin{enumerate}
    \item the ability to act and react quickly and easily;
    \item adherence to the principles laid out in the agile manifesto~\cite{cohen};
    \item the extent to which a team implements the practices associated with one or more ostensibly agile methods.
\end{enumerate}

Some agilists believe that agile principles must be implemented using specific agile practices tailored to their own context~\cite{williams}. However, instead of providing empirical evidence that these practices increase a team's ability to act quickly and easily, proponents of agile simply redefined \textit{agility} as adhering to their methods~\cite{wufka2015explaining}.   

Even if we accept that adopting ostensibly agile practices increases a team's agility, neither the agile manifesto nor popular books espousing agile practices~\cite[e.g][]{beck2000extreme,schwaber} address leadership in-depth or explain how what type of leadership behavior is needed over time. Scrum, in particular, promotes a facilitating role---the ``Scrum Master''---and does not acknowledge the necessity of adapting one's leadership style to one's social context~\cite{schwaber2020ScrumGuide}. In reality, many companies use a mixture of methods and adopt agile methods incrementally~\cite{korhonen2013evaluating}. 

Most leadership research in software engineering views leadership as an individual role~\cite[e.g.][]{kalliamvakou2017makes,garcia2019leadership}). Studies also show how some agile roles include leadership behavior, e.g., Scrum Masters~\cite{srivastava2017leadership} and agile coaches~\cite{backlander2019doing}. 

Some studies investigate specific leadership styles. For example, Veiseh et al.~\cite{veiseh2014study} showed that agile leaders are more transformational. Garcia et al.~\cite{garcia2019leadership} found that the effectiveness of different leadership styles are independent of the software development process (e.g. agile, waterfall). Other studies have investigated how leadership styles change as teams mature~\cite{grenjss2,gren2019agile}. What leadership behavior is needed depends on individual readiness~\cite{hersey} and the group's development stage~\cite{wheelan}. Leadership roles typically become more distributed as a team matures~\cite{spiegler2021empirical,spiegler2020quantitative}. 

Andrias et al.~\cite{andrias2018towards} found three leadership styles---Tigers, Cranes, and Elephants---in Information Systems Development projects. Przybilla et al.~\cite{przybilla2019emergent,przybilla2020conceptual} suggests a research plan and a theoretical model where they want to investigate individual traits (represented by personality traits and cognitive ability) and the perceived leadership distribution to explain self-organization. They hypothesize that many individuals would be perceived as leaders in agile teams. 
Meanwhile, Moe et al.~\cite{moe2009understanding} argue that agile teams should be trained in leadership. The same authors later studied challenges of shared decision-making in agile teams, including ``alignment of strategic product plans with iteration plans, allocation of development resources, and performing development and maintenance tasks in teams''~\cite{moe2012challenges}.

We did not find any software engineering studies that investigate social identity (see Section~\ref{psylead}) and leadership together. However, developers' social identities play a key role in the effectiveness of software projects~\cite{backevik2019social}.

In contrast, much research investigates agile teams' self-organization and self-management~\cite[e.g.][]{moe2008understanding,hoda2012self,ringstad2011agile,hoda2013self}. Furthermore, Kakar~\cite{kakar2017assessing} showed that agile teams have higher degrees of self-organization than teams in a more traditional team setup. We extends these works by differentiating leadership from management and organization. Only one study has investigated explicit leadership aspects of getting teams to self-organize~\cite{gren2020agile}, but from the perspective of an appointed agile leader. 

As early as 2003, scholars argued for including cultural aspects in software process improvement efforts~\cite{ngwenyama2003competing}. Later studies confirmed that considering culture improves outcomes in software process improvement~\cite{shih2010exploring}. Others argued that social norms, values and beliefs are all important when deploying systems development methodologies in general~\cite{iivari2007relationship} and agile software development in particular~\cite{strode2009impact}. Our theme of participants balancing conflicting cultures resonates with these prior studies. 

\subsection{Research on Leadership in Social Psychology}\label{psylead}
Early (19\textsuperscript{th} century) conceptions of leadership were dominated by the \emph{Great Man Theory}~\cite{hogg2003social}.\footnote{We preserve the original, gender-specific name here because it highlights how out-dated this theory is.} It posits that the success of nations, organizations, and endeavors is largely determined by leaders---amazing individuals who changed the world using inherent talent for leading. This theory is no longer taken seriously because both of its premises are false: leadership is a learnable set of skills rather than an innate talent~\cite{chemers2000leadership}, and that the fate of nations is determined by multitudinous factors rather than individuals. Criticism of the \textit{Great Man Theory} is not new; in 1869 Tolstoy wrote: ``To elicit the laws of history we must leave aside kings, ministers and generals, and select for study the homogeneous, infinitesimal elements which influence the masses''~\cite[p. 977]{tolstoy}.

Hogg et al.~\cite{hogg2014sp} categorize 20\textsuperscript{th}-century leadership research into seven different focus areas: 

(1) \emph{Personality traits and individual differences} (including the Great Man Theory): a body of research that views leadership as ``innate or acquired individual characteristics''~\cite[p. 316]{hogg2014sp}.

(2) \emph{Situational perspectives}: a body of research that views leadership as a product of a context, which is considered oversimplified because the characteristics of individual leaders can also matter~\cite{hogg2014sp}. 

(3) \emph{What leaders do}: a body of research that investigates the actual behaviors of leaders and typically divides leaders into task-focused or relationship-focused~\cite{hogg2014sp}. This is problematic because it ignores the interplay between leader behavior and context. 

(4) \emph{Contingency theories}: a body of research that connects situational and behavioral factors. The main problem with these theories is that they are rather static~\cite{hogg2014sp} and only focus on one specific leadership behavior in one specific context at a time. 

(5) \emph{Transactional leadership}: a body of research that focuses on ``the transaction of resources between the leader and the followers''~\cite{de2005and}. These ideas were extended to the leader-member exchange (LMX) theory~\cite{graen1995relationship}, which posits that leader-follower groups proceed through a series of phases similar to Tuckman's stages of group development~\cite{tuckman}. Transactional leadership theories are criticized for failing to appreciate that leadership is a group process~\cite{hogg2005effective}.  

(6) \emph{Transformational leadership}: a body of research that focuses on how leaders can transform goals into action using charisma. This work posits that leaders should offer employees an identity within the organization such that employees take personal responsibility for the success of the company and see company failures as their own~\cite{bass2006transformational}. Transformational leadership improves follower performance at the individual, team and organizational levels~\cite{wang2011transformational}. However, immoral charismatic leaders may use power only for personal gain or to promote their own personal vision, censure critical or opposing views, demand unquestioning acceptance of their decisions, use one-way communicatio, and ignore followers' needs~\cite{howell1992ethics}.

(7) \emph{Leader perceptions}: a body of research built on \textit{social cognition}, a branch of cognitive psychology that focuses on how people process, store, and apply information about other people and social situations~\cite{hogg2014sp}. For instance, Leader Categorization Theory posits that our perception of leadership influences our selection and endorsement of leaders~\cite{lord2001contextual}. 

In summary, the first three areas do not have much explanatory power~\cite{chemers2000leadership}. Transactional leadership has merit, while transformational leadership is quite effective but open to abuse. Understanding perceptions of leadership also helps to explain who people choose to endorse as leaders. 

Most of the leadership studies in software engineering fit into one or more of Hogg et al.'s categories; for example, leadership as an individual role~\cite{kalliamvakou2017makes,garcia2019leadership}, or the style of being a transformational leader~\cite{veiseh2014study}. However, leadership research in the 21\textsuperscript{st} century has taken new directions. As we describe next, leadership is now seen as dynamic and integrated into the social process of group formation. 

\subsection{Dynamic Team Leadership}
Recent studies of team effectiveness suggest that appointed leaders of complex projects create, maintain, and refine collective systems of meaning and expectation, which, in turn, guide the behavior of individuals~\cite{kozlowski2006enhancing}. This different perspective of leadership entails less control and direction and more of the leader's ``developmental and instructional capabilities applied over time''~\cite{kozlowski2006enhancing}. 

Members of effective teams share leadership dynamically~\cite{2009tei}; that is, the distribution of leadership work among team members is constantly in flux. The team constantly adjusts (based on their circumstances) what leadership work is being done, who is doing it and how they are doing it. Ford~\cite{ford2010complex} suggests replacing the term \emph{leadership} with \emph{leadingship} to capture the fact that it is a shared property of the team rather than a property of designated ``leaders''. 

These ideas have existed in parallel with some of the previous leadership theories described above---for example, Cranach~\cite{von1986leadership} suggested that leadership is a function of group action in \citeyear{von1986leadership}. 

In summary, leadership is a kind of work---a shared function of driving the work forward toward the team's goal---which teams share. Teams constantly adjust their shared leadership work to their changing circumstances.

\subsection{Social Identity Theory}
Above, we described Hogg et al.'s seven \textit{historical} areas of leadership research. Hogg et al.'s point, in categorizing leadership research thusly, is to contrast these historical views with a contemporary, \textit{Social Identity Theory} of leadership~\cite{hogg2003social}. Whereas all seven historical areas assume that leadership is an individual \textit{role}; Social Identity Theory models leadership as a \textit{relationship} where group members can influence each other to adopt (and perceive as their own) new values, attitudes and goals. 

A person's \textit{social identity} is the part of their sense of self that derives from group membership; that is their self-categorization into different social groups~\cite{hornsey2008social}. Effective leadership transforms individual action into group action that defines group membership. Leadership is an \textit{identity function}: it consolidates, stabilizes, and anchors the group's identity~\cite{hogg2003social}. For example, the congruence between the social identities of leaders and followers leads to higher ratings of transformational leadership~\cite{avolio2020leader}.

Social identities also play an important role in how people collaborate. To collaborate effectively, teams need to create a common social category. Team members do \textit{not} have to abandon their prior social categories to identify with the team; rather, they create a new social category for the cross-functional team~\cite{dovidio1998intergroup}. For example, if some members of a software development team are also part of group that promotes DevOps across the organization; they need not suppress that part of their social identity to feel like part of the software team as well. Furthermore, the new social category does not require commonalities beyond team membership to take hold. For example, the team members can all have different hobbies, nationalities, belief systems or interests.

However, the more a person identifies with one social category---be it a role or a team---the more they exhibit inter-group bias; that is, favoring the groups we are part of (the ``in-group'') over others (``the out-group'')~\cite{hewstone2002intergroup}. Inter-group bias has both positive and negative effects; for example, inter-group bias can magnify team member's motivation, but it can also lead employees to put their team's best interests ahead of the organization's~\cite{hewstone2002intergroup}.

\subsection{Organizational Culture}
Organizational culture refers to ``the pattern of beliefs, values and learned ways of coping with experience that have developed during the course of an organization’s history, and which tend to be manifested in its material arrangements and in the behaviours of its members''~\cite{brown1998organisational}.
Culture is widely-studied in many fields including anthropology, management, political science, and sociology. For the purposes of this paper, however, the key point is that leadership (and social identity) are inseparable from culture~\cite{scheincultlead}. 

Organizational culture is ``a particularistic system of symbols shaped by ambient society and the organization’s history, leadership and contingencies, differentially shared, used and modified by actors in the course of acting and making sense out of organizational events''~\cite[p. 216]{allaire1984theories}. Rather than a single, monolithic culture, organizations can have a mishmash of conflicting cultural views~\cite{gregory1983native}. Any nuanced understanding of leadership in agile teams therefore must be situated in the organization's ongoing cultural conflict.

\subsection{Summary}
In summary, leadership research has evolved considerably since the 19\textsuperscript{th} century. Three recent trends in leadership research help explain leadership in agile software development: 
\begin{itemize}
    \item social identity helps explain how leaders get individuals to feel like part of a team and accept responsibility;
    \item dynamic team leadership helps explain how leadership work is shared among team members;
    \item the idea that leadership is inextricable from culture helps explain how leaders balance conflicting cultures associated with agile and milestone-driven processes.
\end{itemize}    

More broadly, increased focus on people and their interactions is a core dimension of transitioning to, and maintaining, agile software development. Leadership is part of this increased focus on human factors. Next, we will further integrate our findings back into recent leadership research.

\section{Discussion}\label{sec:discussion}
The main contributions of this paper---the concepts illustrated by Figure~\ref{fig_sim}---resonate well with recent studies on leadership in social psychology, most of which have not previously been applied to the software engineering context. 

\subsection{Dynamic Team Leadership}
Our findings suggest that leadership in cross-functional agile teams is dynamically shared: many team members accept responsibility, take initiative, do leadership work; and furthermore, who does what leadership work changes over time. This resonates with research on dynamic team leadership, which has found that effective leaders develop and guide team members instead of controlling or directing them~\cite{kozlowski2006enhancing}. Agile teams appear to need leadership that consistently adapts to their evolving situation; agile team members need to actively drive teamwork. 

Our results suggest that self-appointed agile leaders let other team members take on leadership work when they (the others) are ready to do so. Not allowing team members to take on leadership work and drive the teamwork forward appears to inhibit the development of the desired agile team dynamic~\cite{2009tei}. Having traditional managers in or around teams can therefore prevent them from becoming agile. 

Agile methods frequently prescribe self-managed teams run without interference from external leaders. However, both our findings and existing research on dynamic team leadership suggest that external or appointed leaders creating, maintaining, and refining expectations for the team can be beneficial~\cite{kozlowski2006enhancing}.

\subsection{Social Identity}
Our findings suggest that building a strong social identity across functions, such that employees identify with their new team in addition to their old role, is important for effective agile leadership. Initially, participants added to their colleagues' identities by simply designing cross-functional teams. Having team members from different functions collaborating was described as one key to breaking the silos and focusing on increased customer value. 

Our results are consistent with prior findings in psychology that designing teams is a kind of leadership behavior~\cite{hogg2003social}. The positive effects of cross-functional teams are also well known and include increased organizational learning, job satisfaction, and overall effectiveness~\cite{denison1996chimneys}. The agile leaders we interviewed described creating a sense of belonging and shared responsibility not only within teams but also with customers and other organizational functions. 

Our interviewees' emphasis on building a strong team identity (i.e.\ clearly defining a social category for the team) is consistent with the main premise of social identity theory: our membership in groups is a major part of who we are and how we fit into our social world. Our interviewees' descriptions of how employees identify with different roles and teams are consistent with the idea that an individual's social identity comprises many social categories~\cite{hewstone2002intergroup,hornsey2008social}. 

Furthermore, synthesizing prior literature with our findings suggests the following recommendation. New team members do not have to abandon their prior social categories to identify with their new team. Instead of trying to remove a person's prior social category (e.g. a software tester), we simply add the cross-functional team identity (e.g. ``I'm a member of team Eta''). The different categories do not interfere with each other, and trying to suppress aspects of a person's social identity is problematic. Moreover, developing employees' social identities can decrease inter-group bias and make collaboration more effective across the organization~\cite[e.g.][]{denison1996chimneys}. 

\subsection{Organizational Culture}
Like prior software engineering research on culture~\cite{iivari,tolfo2008}, our findings suggest that transitioning to agile software development is inextricable from sifting organizational cultures. Realizing the benefits of agile software development requires more than just implementing agile practices and process improvements---the organizational culture has to change from one of individual leaders commanding and controlling to one of decentralized, dynamic team leadership. That is why our participants focused on implementing agile principles, tailored to their specific contexts, rather than inflexibly rolling out a standard set of agile practices.  

Furthermore, the agile leaders saw themselves as shields between the new agile cultures of their teams and the old, milestone and function-oriented cultures in their surrounding company ecosystems. This shielding resonates with prior psychological literature on how understanding organizational culture is essential to making sense of the organization's history, its leadership, and events~\cite{allaire1984theories}. Even the participants from the smaller companies contrasted effective leadership in agile teams to other leadership behavior. 

Consistent with prior research on agile transitions~\cite{iivari,tolfo2008}, our results show that agile leaders perform a cultural balancing act across the organizational functions to facilitate a transition toward agile processes. More general research on organizational culture similarly shows that a multitude of organizational cultures existing simultaneously within large organizations. Multicultural models are needed because large organizations have problems of ``cross-cultural'' conflict~\cite{gregory1983native}. When a company adopts agile, the new agile culture must therefore be balanced with existing, and sometimes conflicting, cultures. For example, some of our participants described adapted to their customers' culture by decreasing their release frequency.

\subsection{Answering our Research Question}

This study sought to address the question, what is the nature of leadership in agile software development, from the perspective of professionals who identify as agile leaders? Put simply, our answer is as follows.

\smallskip
{\narrower \noindent \textit{Agile leadership is dynamically shared among team members, engenders a sense of belonging to the cross-functional team, is inextricable from organizational culture, and often requires balancing competing organizational cultures.}  \par}
\smallskip

\subsection{Implications} \label{sec:implications}
The above findings have several implications for researchers, professionals and educators. 

Future research on leadership in software engineering should be more connected to recent research in social psychology and organizational behavior. ``What makes a good agile leader?'' is the wrong question because effective agile teams \textit{share} leadership work. Counter-intuitively, to understand ``what to look for in a great software engineering manager, and how to further develop their skills to support the teams they manage''~\cite{kalliamvakou2017makes}, we need to conceptualize leadership as a property of a team, rather than an individual. Once we understand that leadership is a kind of work, rather than a kind of person, we can recognize that the effectiveness of software engineering managers depends on their ability to (1) get employees to identify as part of team, (2) get team members to accept greater responsibility, and (3) manage cultural conflict. Indeed, transitioning to agile development may increase the importance of these aspects of leadership.


The proposed model explains that, as organizations gradually shift to agile, they must balance competing cultures, including conflicts between teams, within teams, and even within individuals. Practitioners may benefit from collaborating with academics on this because theories of leadership are challenging to digest and apply. 

Meanwhile, educators may find the proposed model useful for teaching agile project management---a problematic subject for many software engineering educators~\cite[cf.][]{ralph2018re}---especially at the graduate level. The model can be used both to explain to students the relationship between management and leadership in self-managed, self-organized software teams and to frame class discussion of taking on leadership work and responsibilities.

\section{Limitations}\label{vts}
Being purely qualitative, we examine this study's limitations through common qualitative categories including Credibility, Resonance, Originality, Usefulness and Transferability~\cite{charmaz2014constructing,guba1985naturalstic}. 

Credibility is the degree to which findings are demonstrably grounded in observation. We provide detailed analysis tables (see \nameref{suppl} showing how each theme and category is grounded in specific interviewee statements. 

Resonance is the degree to which findings make sense to participants. We assessed resonance using \textit{dialogical interviewing}, a technique in which, during interviews, the interviewer reformulates the interviewee's statements in the interviewer's words to synchronize their interpretations and co-construct meaning~\cite{harvey2015beyond}. This technique was used throughout the interviews as interviewee expressed something that seemed important to them. It lead to many small improvements and revisions to our understanding of the interviewees' perspectives, which filter into the emerging themes in many small ways. 

Originality is the degree to which the study produces novel concepts. While the themes that emerge from this study map into existing theories in social psychology, this is the first study to show that leadership is present in, and critical for, self-managed agile teams. Although dynamic team leadership and social identity are widely researched in social psychology, this is the first study that demonstrates that these theories apply to, and are useful for, understanding software teams. 

Usefulness is the degree to which a study's results can be applied by someone (e.g.\ practitioners, researchers). Section \ref{sec:implications} describes the potential uses of our findings by researchers, professionals and educators. The conceptual transition from leadership as a role to leadership as a shareable form of work has the potential to spawn a new direction in research, education, and practice. 

Transferability is the degree to which other researchers could apply a study's proposed concepts in different contexts. Transferability is an alternative to generalizability for qualitative studies that do not statistically generalize from a sample to a population. 
The mapping of our findings into prior theories, which apply across diverse contexts, suggests high transferability. However, agile leaders are scarcer than developers and much scarcer than consumers, so we were not able to find as many participants as we would have liked. We attempted to mitigate bias among participants by recruiting a diverse group in terms of gender, race, nationality, job description, company size, and industry. Future research could improve confidence in the transferability of results by investigating agile leadership from the perspective of professionals who do \textit{not} identify as agile leaders, and professionals from a wider variety of organizational contexts.

Moreover, the analysis may be colored by the first author's experiences with agile leadership at Volvo Cars; specifically, the first author's practical experience with agile culture clashes may have affected the coding of the data. We attempted to mitigate this threat by having the second author audit the coding. Similarly, both of the authors have backgrounds in social psychology, which could affect the degree to which our findings converge with existing psychological theory in ways we can neither perceive nor mitigate.

\section{Conclusions and Future Work}\label{con}
This study was motivated by the three realizations: (1) leadership is crucial for projects and organizations, (2) little research has investigated how leadership works in self-managed agile teams, and (3) software engineering research has not embraced recent advances in psychological research on leadership. We therefore conducted a qualitative survey of thirteen software professionals who identify as agile leaders. Our main contribution is the model of agile leadership shown in Figure~\ref{fig_sim}. Briefly, the model suggests that: (1) leadership is a kind of work rather than a job role; (2) self-organized\slash self-managed teams dynamically share leadership work among team members; (3) employees are more willing to accept leadership work and responsibility when they identify as belonging to their teams; and (4) effective leadership often involves balancing conflicting cultures (e.g. agile vs. milestone-driven).  

This qualitative study aims to explain rather than predict. 
Future work could quantitatively assess the antecedents and consequences of effective leadership using validated scales for concepts including social identity~\cite{luhtanen1992collective}, or explore under-researched aspects of leadership such as \textit{prototypicality} (the idea that more typical team members are better at leadership work~\cite{hogg2003social}) and inter-group bias (favoritism toward one's teammates~\cite{hewstone2002intergroup}) in software teams.  

Software engineering researchers need to get better at integrating theories and findings from reference disciplines and collaborating directly with scientists from other disciplines. While cross-disciplinary research is intrinsically challenging, software engineering is intrinsically interdisciplinary. Therefore, only by appreciating research from other disciplines can we effectively understand, explain, predict and prescribe software engineering phenomena.

\section*{Data Availability}\label{suppl}

Supplementary materials including our interview guide, descriptions of participants' organizations, and analysis tables are available at \url{https://doi.org/10.5281/zenodo.5816890}. Interview transcripts have not been published to mitigate re-identification risk.

\section*{Acknowledgments}
We would like to thank our participants for investing their time and effort to make this work possible, and Alfredo Goldman and Richard Torkar for their support.

\bibliographystyle{ACM-Reference-Format}
\bibliography{references}


\begin{thebibliography}{00}


\ifx \showCODEN    \undefined \def \showCODEN     #1{\unskip}     \fi
\ifx \showDOI      \undefined \def \showDOI       #1{{\tt DOI:}\penalty0{#1}\ }
  \fi
\ifx \showISBNx    \undefined \def \showISBNx     #1{\unskip}     \fi
\ifx \showISBNxiii \undefined \def \showISBNxiii  #1{\unskip}     \fi
\ifx \showISSN     \undefined \def \showISSN      #1{\unskip}     \fi
\ifx \showLCCN     \undefined \def \showLCCN      #1{\unskip}     \fi
\ifx \shownote     \undefined \def \shownote      #1{#1}          \fi
\ifx \showarticletitle \undefined \def \showarticletitle #1{#1}   \fi
\ifx \showURL      \undefined \def \showURL       #1{#1}          \fi
\providecommand\bibfield[2]{#2}
\providecommand\bibinfo[2]{#2}
\providecommand\natexlab[1]{#1}
\providecommand\showeprint[2][]{arXiv:#2}

\bibitem[\protect\citeauthoryear{Aldiabat and Le~Navenec}{Aldiabat and
  Le~Navenec}{2018}]%
        {aldiabat2018data}
\bibfield{author}{\bibinfo{person}{Khaldoun~M Aldiabat} {and}
  \bibinfo{person}{Carole-Lynne Le~Navenec}.} \bibinfo{year}{2018}\natexlab{}.
\newblock \showarticletitle{Data saturation: The mysterious step in grounded
  theory methodology}.
\newblock \bibinfo{journal}{{\em The Qualitative Report\/}}
  \bibinfo{volume}{23}, \bibinfo{number}{1} (\bibinfo{year}{2018}),
  \bibinfo{pages}{245--261}.
\newblock


\bibitem[\protect\citeauthoryear{Allaire and Firsirotu}{Allaire and
  Firsirotu}{1984}]%
        {allaire1984theories}
\bibfield{author}{\bibinfo{person}{Yvan Allaire} {and}
  \bibinfo{person}{Mihaela~E Firsirotu}.} \bibinfo{year}{1984}\natexlab{}.
\newblock \showarticletitle{Theories of organizational culture}.
\newblock \bibinfo{journal}{{\em Organization studies\/}} \bibinfo{volume}{5},
  \bibinfo{number}{3} (\bibinfo{year}{1984}), \bibinfo{pages}{193--226}.
\newblock


\bibitem[\protect\citeauthoryear{Andrias, Matook, and Vidgen}{Andrias
  et~al\mbox{.}}{2018}]%
        {andrias2018towards}
\bibfield{author}{\bibinfo{person}{Mone~Stepanus Andrias},
  \bibinfo{person}{Sabine Matook}, {and} \bibinfo{person}{Richard Vidgen}.}
  \bibinfo{year}{2018}\natexlab{}.
\newblock \showarticletitle{Towards a typology of agile {ISD} leadership}. In
  \bibinfo{booktitle}{{\em Research-in-Progress Papers, Proceedings of the
  European Conference on Information Systems}}, Vol.~\bibinfo{volume}{28}. AIS.
\newblock


\bibitem[\protect\citeauthoryear{Avolio, Keng-Highberger, Lord, Hannah,
  Schaubroeck, and Kozlowski}{Avolio et~al\mbox{.}}{2020}]%
        {avolio2020leader}
\bibfield{author}{\bibinfo{person}{Bruce~J Avolio}, \bibinfo{person}{Fong~T
  Keng-Highberger}, \bibinfo{person}{Robert~G Lord}, \bibinfo{person}{Sean~T
  Hannah}, \bibinfo{person}{John~M Schaubroeck}, {and}
  \bibinfo{person}{Steve~WJ Kozlowski}.} \bibinfo{year}{2020}\natexlab{}.
\newblock \showarticletitle{How leader and follower prototypical and
  antitypical attributes influence ratings of transformational leadership in an
  extreme context}.
\newblock \bibinfo{journal}{{\em Human Relations\/}} (\bibinfo{year}{2020}).
\newblock
\showDOI{%
\url{http://dx.doi.org/10.1177/0018726720958040}}


\bibitem[\protect\citeauthoryear{B{\"a}ckevik, Thol{\'e}n, and
  Gren}{B{\"a}ckevik et~al\mbox{.}}{2019}]%
        {backevik2019social}
\bibfield{author}{\bibinfo{person}{Andreas B{\"a}ckevik}, \bibinfo{person}{Erik
  Thol{\'e}n}, {and} \bibinfo{person}{Lucas Gren}.}
  \bibinfo{year}{2019}\natexlab{}.
\newblock \showarticletitle{Social identity in software development}. In
  \bibinfo{booktitle}{{\em Proceedings of the 12th International Workshop on
  Cooperative and Human Aspects of Software Engineering}}.
  \bibinfo{publisher}{IEEE Press}, \bibinfo{pages}{107--114}.
\newblock


\bibitem[\protect\citeauthoryear{B{\"a}cklander}{B{\"a}cklander}{2019}]%
        {backlander2019doing}
\bibfield{author}{\bibinfo{person}{Gisela B{\"a}cklander}.}
  \bibinfo{year}{2019}\natexlab{}.
\newblock \showarticletitle{Doing complexity leadership theory: {H}ow agile
  coaches at Spotify practise enabling leadership}.
\newblock \bibinfo{journal}{{\em Creativity and Innovation Management\/}}
  \bibinfo{volume}{28}, \bibinfo{number}{1} (\bibinfo{year}{2019}),
  \bibinfo{pages}{42--60}.
\newblock


\bibitem[\protect\citeauthoryear{Baltes and Ralph}{Baltes and Ralph}{ress}]%
        {baltes2020sampling}
\bibfield{author}{\bibinfo{person}{Sebastian Baltes} {and}
  \bibinfo{person}{Paul Ralph}.} \bibinfo{year}{in press}\natexlab{}.
\newblock \showarticletitle{Sampling in software engineering research: A
  critical review and guidelines}.
\newblock \bibinfo{journal}{{\em Empirical Software Engineering\/}}
  (\bibinfo{year}{in press}).
\newblock


\bibitem[\protect\citeauthoryear{Bass and Riggio}{Bass and Riggio}{2006}]%
        {bass2006transformational}
\bibfield{author}{\bibinfo{person}{Bernard~M Bass} {and}
  \bibinfo{person}{Ronald~E Riggio}.} \bibinfo{year}{2006}\natexlab{}.
\newblock \bibinfo{booktitle}{{\em Transformational leadership}}.
\newblock \bibinfo{publisher}{Psychology press}.
\newblock


\bibitem[\protect\citeauthoryear{Beck}{Beck}{1999}]%
        {beck1999embracing}
\bibfield{author}{\bibinfo{person}{Kent Beck}.}
  \bibinfo{year}{1999}\natexlab{}.
\newblock \showarticletitle{Embracing change with extreme programming}.
\newblock \bibinfo{journal}{{\em IEEE Computer\/}} \bibinfo{volume}{32},
  \bibinfo{number}{10} (\bibinfo{year}{1999}), \bibinfo{pages}{70--77}.
\newblock


\bibitem[\protect\citeauthoryear{Beck}{Beck}{2000}]%
        {beck2000extreme}
\bibfield{author}{\bibinfo{person}{Kent Beck}.}
  \bibinfo{year}{2000}\natexlab{}.
\newblock \bibinfo{booktitle}{{\em Extreme programming explained: embrace
  change}}.
\newblock \bibinfo{publisher}{Addison-Wesley Professional}.
\newblock


\bibitem[\protect\citeauthoryear{Braun and Clarke}{Braun and Clarke}{2006}]%
        {braun2006using}
\bibfield{author}{\bibinfo{person}{Virginia Braun} {and}
  \bibinfo{person}{Victoria Clarke}.} \bibinfo{year}{2006}\natexlab{}.
\newblock \showarticletitle{Using thematic analysis in psychology}.
\newblock \bibinfo{journal}{{\em Qualitative research in psychology\/}}
  \bibinfo{volume}{3}, \bibinfo{number}{2} (\bibinfo{year}{2006}),
  \bibinfo{pages}{77--101}.
\newblock


\bibitem[\protect\citeauthoryear{Brown}{Brown}{1998}]%
        {brown1998organisational}
\bibfield{author}{\bibinfo{person}{A.D. Brown}.}
  \bibinfo{year}{1998}\natexlab{}.
\newblock \bibinfo{booktitle}{{\em Organisational Culture\/}
  (\bibinfo{edition}{2nd} ed.)}.
\newblock \bibinfo{publisher}{Financial Times}.
\newblock
\showISBNx{9780273631477}


\bibitem[\protect\citeauthoryear{Charmaz}{Charmaz}{2014}]%
        {charmaz2014constructing}
\bibfield{author}{\bibinfo{person}{Kathy Charmaz}.}
  \bibinfo{year}{2014}\natexlab{}.
\newblock \bibinfo{booktitle}{{\em Constructing grounded theory}}.
\newblock \bibinfo{publisher}{sage}.
\newblock


\bibitem[\protect\citeauthoryear{Chemers}{Chemers}{2000}]%
        {chemers2000leadership}
\bibfield{author}{\bibinfo{person}{Martin~M Chemers}.}
  \bibinfo{year}{2000}\natexlab{}.
\newblock \showarticletitle{Leadership research and theory: {A} functional
  integration}.
\newblock \bibinfo{journal}{{\em Group Dynamics: Theory, research, and
  practice\/}} \bibinfo{volume}{4}, \bibinfo{number}{1} (\bibinfo{year}{2000}),
  \bibinfo{pages}{27}.
\newblock


\bibitem[\protect\citeauthoryear{Cohen, Lindvall, and Costa}{Cohen
  et~al\mbox{.}}{2004}]%
        {cohen}
\bibfield{author}{\bibinfo{person}{David Cohen}, \bibinfo{person}{Mikael
  Lindvall}, {and} \bibinfo{person}{Patricia Costa}.}
  \bibinfo{year}{2004}\natexlab{}.
\newblock \showarticletitle{An introduction to agile methods}.
\newblock \bibinfo{journal}{{\em Advances in Computers\/}}
  \bibinfo{volume}{62} (\bibinfo{year}{2004}), \bibinfo{pages}{1--66}.
\newblock


\bibitem[\protect\citeauthoryear{De~Cremer and Van~Dijk}{De~Cremer and
  Van~Dijk}{2005}]%
        {de2005and}
\bibfield{author}{\bibinfo{person}{David De~Cremer} {and} \bibinfo{person}{Eric
  Van~Dijk}.} \bibinfo{year}{2005}\natexlab{}.
\newblock \showarticletitle{When and why leaders put themselves first: Leader
  behaviour in resource allocations as a function of feeling entitled}.
\newblock \bibinfo{journal}{{\em European Journal of Social Psychology\/}}
  \bibinfo{volume}{35}, \bibinfo{number}{4} (\bibinfo{year}{2005}),
  \bibinfo{pages}{553--563}.
\newblock


\bibitem[\protect\citeauthoryear{Denison, Hart, and Kahn}{Denison
  et~al\mbox{.}}{1996}]%
        {denison1996chimneys}
\bibfield{author}{\bibinfo{person}{Daniel~R Denison}, \bibinfo{person}{Stuart~L
  Hart}, {and} \bibinfo{person}{Joel~A Kahn}.} \bibinfo{year}{1996}\natexlab{}.
\newblock \showarticletitle{From chimneys to cross-functional teams:
  {D}eveloping and validating a diagnostic model}.
\newblock \bibinfo{journal}{{\em Academy of management journal\/}}
  \bibinfo{volume}{39}, \bibinfo{number}{4} (\bibinfo{year}{1996}),
  \bibinfo{pages}{1005--1023}.
\newblock


\bibitem[\protect\citeauthoryear{Dovidio, Gaertner, and Validzic}{Dovidio
  et~al\mbox{.}}{1998}]%
        {dovidio1998intergroup}
\bibfield{author}{\bibinfo{person}{John~F Dovidio}, \bibinfo{person}{Samuel~L
  Gaertner}, {and} \bibinfo{person}{Ana Validzic}.}
  \bibinfo{year}{1998}\natexlab{}.
\newblock \showarticletitle{Intergroup bias: status, differentiation, and a
  common in-group identity.}
\newblock \bibinfo{journal}{{\em Journal of personality and social
  psychology\/}} \bibinfo{volume}{75}, \bibinfo{number}{1}
  (\bibinfo{year}{1998}), \bibinfo{pages}{109--120}.
\newblock


\bibitem[\protect\citeauthoryear{Felps, Mitchell, and Byington}{Felps
  et~al\mbox{.}}{2006}]%
        {felps2006and}
\bibfield{author}{\bibinfo{person}{Will Felps}, \bibinfo{person}{Terence~R
  Mitchell}, {and} \bibinfo{person}{Eliza Byington}.}
  \bibinfo{year}{2006}\natexlab{}.
\newblock \showarticletitle{How, when, and why bad apples spoil the barrel:
  {N}egative group members and dysfunctional groups}.
\newblock \bibinfo{journal}{{\em Research in organizational behavior\/}}
  \bibinfo{volume}{27} (\bibinfo{year}{2006}), \bibinfo{pages}{175--222}.
\newblock


\bibitem[\protect\citeauthoryear{Ford}{Ford}{2010}]%
        {ford2010complex}
\bibfield{author}{\bibinfo{person}{Randal Ford}.}
  \bibinfo{year}{2010}\natexlab{}.
\newblock \showarticletitle{Complex adaptive leading-ship and open-processional
  change processes}.
\newblock \bibinfo{journal}{{\em Leadership \& Organization Development
  Journal\/}} \bibinfo{volume}{31}, \bibinfo{number}{5} (\bibinfo{year}{2010}),
  \bibinfo{pages}{420--435}.
\newblock


\bibitem[\protect\citeauthoryear{Fowler and Highsmith}{Fowler and
  Highsmith}{2001}]%
        {fowler2001}
\bibfield{author}{\bibinfo{person}{M. Fowler} {and} \bibinfo{person}{J.
  Highsmith}.} \bibinfo{year}{2001}\natexlab{}.
\newblock \bibinfo{title}{{The Agile Manifesto}}.
\newblock \bibinfo{howpublished}{In Software Development, Issue on Agile
  Methodologies, last accessed on December 29th, 2006}.   (\bibinfo{date}{Aug.}
  \bibinfo{year}{2001}).
\newblock


\bibitem[\protect\citeauthoryear{Garcia and Russoi}{Garcia and Russoi}{2019}]%
        {garcia2019leadership}
\bibfield{author}{\bibinfo{person}{Fernando Andre~Zemuner Garcia} {and}
  \bibinfo{person}{Ros{\'a}ria de Fatima Segger~Macri Russoi}.}
  \bibinfo{year}{2019}\natexlab{}.
\newblock \showarticletitle{Leadership and Performance of the Software
  Development Team: Influence of the Type of Project Management}.
\newblock \bibinfo{journal}{{\em Revista Brasileira de Gest{\~a}o de
  Neg{\'o}cios\/}} \bibinfo{volume}{21}, \bibinfo{number}{5}
  (\bibinfo{year}{2019}), \bibinfo{pages}{970--1005}.
\newblock


\bibitem[\protect\citeauthoryear{Graen and Uhl-Bien}{Graen and
  Uhl-Bien}{1995}]%
        {graen1995relationship}
\bibfield{author}{\bibinfo{person}{George~B Graen} {and} \bibinfo{person}{Mary
  Uhl-Bien}.} \bibinfo{year}{1995}\natexlab{}.
\newblock \showarticletitle{Relationship-based approach to leadership:
  {D}evelopment of leader-member exchange (LMX) theory of leadership over 25
  years: {A}pplying a multi-level multi-domain perspective}.
\newblock \bibinfo{journal}{{\em The leadership quarterly\/}}
  \bibinfo{volume}{6}, \bibinfo{number}{2} (\bibinfo{year}{1995}),
  \bibinfo{pages}{219--247}.
\newblock


\bibitem[\protect\citeauthoryear{Gregory}{Gregory}{1983}]%
        {gregory1983native}
\bibfield{author}{\bibinfo{person}{Kathleen~L Gregory}.}
  \bibinfo{year}{1983}\natexlab{}.
\newblock \showarticletitle{Native-view paradigms: Multiple cultures and
  culture conflicts in organizations}.
\newblock \bibinfo{journal}{{\em Administrative science quarterly\/}}
  \bibinfo{volume}{28}, \bibinfo{number}{3} (\bibinfo{year}{1983}),
  \bibinfo{pages}{359--376}.
\newblock


\bibitem[\protect\citeauthoryear{Gren, Goldman, and Jacobsson}{Gren
  et~al\mbox{.}}{2020}]%
        {gren2019agile}
\bibfield{author}{\bibinfo{person}{Lucas Gren}, \bibinfo{person}{Alfredo
  Goldman}, {and} \bibinfo{person}{Christian Jacobsson}.}
  \bibinfo{year}{2020}\natexlab{}.
\newblock \showarticletitle{Agile ways of working: {A} team maturity
  perspective}.
\newblock \bibinfo{journal}{{\em Journal of Software: Evolution and Process\/}}
  \bibinfo{volume}{32}, \bibinfo{number}{6} (\bibinfo{year}{2020}),
  \bibinfo{pages}{e2244}.
\newblock


\bibitem[\protect\citeauthoryear{Gren and Lindman}{Gren and Lindman}{2020}]%
        {gren2020agile}
\bibfield{author}{\bibinfo{person}{Lucas Gren} {and} \bibinfo{person}{Magdalena
  Lindman}.} \bibinfo{year}{2020}\natexlab{}.
\newblock \showarticletitle{What an Agile Leader Does: The Group Dynamics
  Perspective}. In \bibinfo{booktitle}{{\em International Conference on Agile
  Software Development}}. \bibinfo{publisher}{Springer},
  \bibinfo{pages}{178--194}.
\newblock


\bibitem[\protect\citeauthoryear{Gren, Torkar, and Feldt}{Gren
  et~al\mbox{.}}{2017}]%
        {grenjss2}
\bibfield{author}{\bibinfo{person}{L Gren}, \bibinfo{person}{R Torkar}, {and}
  \bibinfo{person}{R Feldt}.} \bibinfo{year}{2017}\natexlab{}.
\newblock \showarticletitle{Group development and group maturity when building
  agile teams: {A} qualitative and quantitative investigation at eight large
  companies}.
\newblock \bibinfo{journal}{{\em The Journal of Systems and Software\/}}
  \bibinfo{volume}{124} (\bibinfo{year}{2017}), \bibinfo{pages}{104—--119}.
\newblock
\showDOI{%
\url{http://dx.doi.org/10.1016/j.jss.2016.11.024}}


\bibitem[\protect\citeauthoryear{Harvey}{Harvey}{2015}]%
        {harvey2015beyond}
\bibfield{author}{\bibinfo{person}{Lou Harvey}.}
  \bibinfo{year}{2015}\natexlab{}.
\newblock \showarticletitle{Beyond member-checking: A dialogic approach to the
  research interview}.
\newblock \bibinfo{journal}{{\em International Journal of Research \& Method in
  Education\/}} \bibinfo{volume}{38}, \bibinfo{number}{1}
  (\bibinfo{year}{2015}), \bibinfo{pages}{23--38}.
\newblock


\bibitem[\protect\citeauthoryear{Hersey, Blanchard, and Natemeyer}{Hersey
  et~al\mbox{.}}{1979}]%
        {hersey}
\bibfield{author}{\bibinfo{person}{Paul Hersey}, \bibinfo{person}{Kenneth~H
  Blanchard}, {and} \bibinfo{person}{Walter~E Natemeyer}.}
  \bibinfo{year}{1979}\natexlab{}.
\newblock \showarticletitle{Situational leadership, perception, and the impact
  of power}.
\newblock \bibinfo{journal}{{\em Group \& Organization Management\/}}
  \bibinfo{volume}{4}, \bibinfo{number}{4} (\bibinfo{year}{1979}),
  \bibinfo{pages}{418--428}.
\newblock


\bibitem[\protect\citeauthoryear{Hewstone, Rubin, and Willis}{Hewstone
  et~al\mbox{.}}{2002}]%
        {hewstone2002intergroup}
\bibfield{author}{\bibinfo{person}{Miles Hewstone}, \bibinfo{person}{Mark
  Rubin}, {and} \bibinfo{person}{Hazel Willis}.}
  \bibinfo{year}{2002}\natexlab{}.
\newblock \showarticletitle{Intergroup bias}.
\newblock \bibinfo{journal}{{\em Annual review of psychology\/}}
  \bibinfo{volume}{53}, \bibinfo{number}{1} (\bibinfo{year}{2002}),
  \bibinfo{pages}{575--604}.
\newblock


\bibitem[\protect\citeauthoryear{Hoda, Noble, and Marshall}{Hoda
  et~al\mbox{.}}{2012}]%
        {hoda2012self}
\bibfield{author}{\bibinfo{person}{Rashina Hoda}, \bibinfo{person}{James
  Noble}, {and} \bibinfo{person}{Stuart Marshall}.}
  \bibinfo{year}{2012}\natexlab{}.
\newblock \showarticletitle{Self-organizing roles on agile software development
  teams}.
\newblock \bibinfo{journal}{{\em IEEE Transactions on Software Engineering\/}}
  \bibinfo{volume}{39}, \bibinfo{number}{3} (\bibinfo{year}{2012}),
  \bibinfo{pages}{422--444}.
\newblock


\bibitem[\protect\citeauthoryear{Hoda, Noble, and Marshall}{Hoda
  et~al\mbox{.}}{2013}]%
        {hoda2013self}
\bibfield{author}{\bibinfo{person}{Rashina Hoda}, \bibinfo{person}{James
  Noble}, {and} \bibinfo{person}{Stuart Marshall}.}
  \bibinfo{year}{2013}\natexlab{}.
\newblock \showarticletitle{Self-organizing roles on agile software development
  teams}.
\newblock \bibinfo{journal}{{\em IEEE Transactions on Software Engineering\/}}
  \bibinfo{volume}{39}, \bibinfo{number}{3} (\bibinfo{year}{2013}),
  \bibinfo{pages}{422--444}.
\newblock


\bibitem[\protect\citeauthoryear{Hogg, Martin, Epitropaki, Mankad, Svensson,
  and Weeden}{Hogg et~al\mbox{.}}{2005}]%
        {hogg2005effective}
\bibfield{author}{\bibinfo{person}{Michael~A Hogg}, \bibinfo{person}{Robin
  Martin}, \bibinfo{person}{Olga Epitropaki}, \bibinfo{person}{Aditi Mankad},
  \bibinfo{person}{Alicia Svensson}, {and} \bibinfo{person}{Karen Weeden}.}
  \bibinfo{year}{2005}\natexlab{}.
\newblock \showarticletitle{Effective leadership in salient groups:
  {R}evisiting leader-member exchange theory from the perspective of the social
  identity theory of leadership}.
\newblock \bibinfo{journal}{{\em Personality and Social Psychology Bulletin\/}}
  \bibinfo{volume}{31}, \bibinfo{number}{7} (\bibinfo{year}{2005}),
  \bibinfo{pages}{991--1004}.
\newblock


\bibitem[\protect\citeauthoryear{Hogg and van Knippenberg}{Hogg and van
  Knippenberg}{2003}]%
        {hogg2003social}
\bibfield{author}{\bibinfo{person}{Michael~A Hogg} {and} \bibinfo{person}{Daan
  van Knippenberg}.} \bibinfo{year}{2003}\natexlab{}.
\newblock \showarticletitle{Social identity and leadership processes in
  groups}.
\newblock \bibinfo{journal}{{\em Advances in Experimental Social Psychology\/}}
   \bibinfo{volume}{35} (\bibinfo{year}{2003}), \bibinfo{pages}{1--52}.
\newblock


\bibitem[\protect\citeauthoryear{Hogg and Vaughan}{Hogg and Vaughan}{2014}]%
        {hogg2014sp}
\bibfield{author}{\bibinfo{person}{Michael~A. Hogg} {and}
  \bibinfo{person}{Graham~M. Vaughan}.} \bibinfo{year}{2014}\natexlab{}.
\newblock \bibinfo{booktitle}{{\em Social Psychology\/} (\bibinfo{edition}{7}
  ed.)}.
\newblock \bibinfo{publisher}{Pearson}, \bibinfo{address}{Harlow, England}.
\newblock


\bibitem[\protect\citeauthoryear{Hornsey}{Hornsey}{2008}]%
        {hornsey2008social}
\bibfield{author}{\bibinfo{person}{Matthew~J Hornsey}.}
  \bibinfo{year}{2008}\natexlab{}.
\newblock \showarticletitle{Social identity theory and self-categorization
  theory: A historical review}.
\newblock \bibinfo{journal}{{\em Social and Personality Psychology Compass\/}}
  \bibinfo{volume}{2}, \bibinfo{number}{1} (\bibinfo{year}{2008}),
  \bibinfo{pages}{204--222}.
\newblock


\bibitem[\protect\citeauthoryear{House and Aditya}{House and Aditya}{1997}]%
        {house1997social}
\bibfield{author}{\bibinfo{person}{Robert~J House} {and} \bibinfo{person}{Ram~N
  Aditya}.} \bibinfo{year}{1997}\natexlab{}.
\newblock \showarticletitle{The social scientific study of leadership: Quo
  vadis?}
\newblock \bibinfo{journal}{{\em Journal of management\/}}
  \bibinfo{volume}{23}, \bibinfo{number}{3} (\bibinfo{year}{1997}),
  \bibinfo{pages}{409--473}.
\newblock


\bibitem[\protect\citeauthoryear{Howell and Avolio}{Howell and Avolio}{1992}]%
        {howell1992ethics}
\bibfield{author}{\bibinfo{person}{Jane~M Howell} {and}
  \bibinfo{person}{Bruce~J Avolio}.} \bibinfo{year}{1992}\natexlab{}.
\newblock \showarticletitle{The ethics of charismatic leadership: submission or
  liberation?}
\newblock \bibinfo{journal}{{\em Academy of Management Perspectives\/}}
  \bibinfo{volume}{6}, \bibinfo{number}{2} (\bibinfo{year}{1992}),
  \bibinfo{pages}{43--54}.
\newblock


\bibitem[\protect\citeauthoryear{Iivari and Huisman}{Iivari and
  Huisman}{2007}]%
        {iivari2007relationship}
\bibfield{author}{\bibinfo{person}{Juhani Iivari} {and} \bibinfo{person}{Magda
  Huisman}.} \bibinfo{year}{2007}\natexlab{}.
\newblock \showarticletitle{The relationship between organizational culture and
  the deployment of systems development methodologies}.
\newblock \bibinfo{journal}{{\em Mis Quarterly\/}} \bibinfo{volume}{31},
  \bibinfo{number}{1} (\bibinfo{year}{2007}), \bibinfo{pages}{35--58}.
\newblock


\bibitem[\protect\citeauthoryear{Iivari and Iivari}{Iivari and Iivari}{2011}]%
        {iivari}
\bibfield{author}{\bibinfo{person}{J Iivari} {and} \bibinfo{person}{N Iivari}.}
  \bibinfo{year}{2011}\natexlab{}.
\newblock \showarticletitle{The relationship between organizational culture and
  the deployment of agile methods}.
\newblock \bibinfo{journal}{{\em Information and Software Technology\/}}
  \bibinfo{volume}{53}, \bibinfo{number}{5} (\bibinfo{year}{2011}),
  \bibinfo{pages}{509--520}.
\newblock


\bibitem[\protect\citeauthoryear{J{\o}rgensen, Mohagheghi, and
  Grimstad}{J{\o}rgensen et~al\mbox{.}}{2017}]%
        {jorgensen2017direct}
\bibfield{author}{\bibinfo{person}{Magne J{\o}rgensen},
  \bibinfo{person}{Parastoo Mohagheghi}, {and} \bibinfo{person}{Stein
  Grimstad}.} \bibinfo{year}{2017}\natexlab{}.
\newblock \showarticletitle{Direct and indirect connections between type of
  contract and software project outcome}.
\newblock \bibinfo{journal}{{\em International Journal of Project
  Management\/}} \bibinfo{volume}{35}, \bibinfo{number}{8}
  (\bibinfo{year}{2017}), \bibinfo{pages}{1573--1586}.
\newblock


\bibitem[\protect\citeauthoryear{Kakar}{Kakar}{2017}]%
        {kakar2017assessing}
\bibfield{author}{\bibinfo{person}{Adarsh~Kumar Kakar}.}
  \bibinfo{year}{2017}\natexlab{}.
\newblock \showarticletitle{Assessing self-organization in Agile software
  development teams}.
\newblock \bibinfo{journal}{{\em Journal of computer information systems\/}}
  \bibinfo{volume}{57}, \bibinfo{number}{3} (\bibinfo{year}{2017}),
  \bibinfo{pages}{208--217}.
\newblock


\bibitem[\protect\citeauthoryear{Kalliamvakou, Bird, Zimmermann, Begel, DeLine,
  and German}{Kalliamvakou et~al\mbox{.}}{2019}]%
        {kalliamvakou2017makes}
\bibfield{author}{\bibinfo{person}{Eirini Kalliamvakou},
  \bibinfo{person}{Christian Bird}, \bibinfo{person}{Thomas Zimmermann},
  \bibinfo{person}{Andrew Begel}, \bibinfo{person}{Robert DeLine}, {and}
  \bibinfo{person}{Daniel~M German}.} \bibinfo{year}{2019}\natexlab{}.
\newblock \showarticletitle{What Makes a Great Manager of Software Engineers?}
\newblock \bibinfo{journal}{{\em IEEE Transactions on Software Engineering\/}}
  \bibinfo{volume}{45}, \bibinfo{number}{1} (\bibinfo{year}{2019}),
  \bibinfo{pages}{87--106}.
\newblock
\showDOI{%
\url{http://dx.doi.org/10.1109/TSE.2017.2768368}}


\bibitem[\protect\citeauthoryear{Korhonen}{Korhonen}{2013}]%
        {korhonen2013evaluating}
\bibfield{author}{\bibinfo{person}{Kirsi Korhonen}.}
  \bibinfo{year}{2013}\natexlab{}.
\newblock \showarticletitle{Evaluating the impact of an agile transformation: a
  longitudinal case study in a distributed context}.
\newblock \bibinfo{journal}{{\em Software Quality Journal\/}}
  \bibinfo{volume}{21}, \bibinfo{number}{4} (\bibinfo{year}{2013}),
  \bibinfo{pages}{599--624}.
\newblock


\bibitem[\protect\citeauthoryear{Kozlowski and Ilgen}{Kozlowski and
  Ilgen}{2006}]%
        {kozlowski2006enhancing}
\bibfield{author}{\bibinfo{person}{Steve~WJ Kozlowski} {and}
  \bibinfo{person}{Daniel~R Ilgen}.} \bibinfo{year}{2006}\natexlab{}.
\newblock \showarticletitle{Enhancing the effectiveness of work groups and
  teams}.
\newblock \bibinfo{journal}{{\em Psychological science in the public
  interest\/}} \bibinfo{volume}{7}, \bibinfo{number}{3} (\bibinfo{year}{2006}),
  \bibinfo{pages}{77--124}.
\newblock


\bibitem[\protect\citeauthoryear{Kozlowski, Watola, Jensen, Kim, and
  Botero}{Kozlowski et~al\mbox{.}}{2009}]%
        {2009tei}
\bibfield{author}{\bibinfo{person}{Steve~WJ Kozlowski},
  \bibinfo{person}{Daniel~J Watola}, \bibinfo{person}{Jaclyn~M Jensen},
  \bibinfo{person}{Brian~H Kim}, {and} \bibinfo{person}{Isabel~C Botero}.}
  \bibinfo{year}{2009}\natexlab{}.
\newblock \showarticletitle{Developing adaptive teams: {A} theory of dynamic
  team leadership}.
\newblock In \bibinfo{booktitle}{{\em Team effectiveness in complex
  organizations: {C}ross-disciplinary perspectives and approaches}},
  \bibfield{editor}{\bibinfo{person}{Eduardo. Salas},
  \bibinfo{person}{Gerald~F. Goodwin}, {and} \bibinfo{person}{C.~Shawn. Burke}}
  (Eds.). \bibinfo{publisher}{Routledge}, \bibinfo{address}{New York, US},
  \bibinfo{pages}{113--155}.
\newblock


\bibitem[\protect\citeauthoryear{Lincoln and Guba}{Lincoln and Guba}{1985}]%
        {guba1985naturalstic}
\bibfield{author}{\bibinfo{person}{Yvonna~S. Lincoln} {and}
  \bibinfo{person}{Egon~G. Guba}.} \bibinfo{year}{1985}\natexlab{}.
\newblock \bibinfo{booktitle}{{\em Naturalistic Inquiry}}.
\newblock \bibinfo{publisher}{Sage}.
\newblock


\bibitem[\protect\citeauthoryear{Lord, Brown, Harvey, and Hall}{Lord
  et~al\mbox{.}}{2001}]%
        {lord2001contextual}
\bibfield{author}{\bibinfo{person}{Robert~G Lord}, \bibinfo{person}{Douglas~J
  Brown}, \bibinfo{person}{Jennifer~L Harvey}, {and} \bibinfo{person}{Rosalie~J
  Hall}.} \bibinfo{year}{2001}\natexlab{}.
\newblock \showarticletitle{Contextual constraints on prototype generation and
  their multilevel consequences for leadership perceptions}.
\newblock \bibinfo{journal}{{\em The Leadership Quarterly\/}}
  \bibinfo{volume}{12}, \bibinfo{number}{3} (\bibinfo{year}{2001}),
  \bibinfo{pages}{311--338}.
\newblock


\bibitem[\protect\citeauthoryear{Luhtanen and Crocker}{Luhtanen and
  Crocker}{1992}]%
        {luhtanen1992collective}
\bibfield{author}{\bibinfo{person}{Riia Luhtanen} {and}
  \bibinfo{person}{Jennifer Crocker}.} \bibinfo{year}{1992}\natexlab{}.
\newblock \showarticletitle{A collective self-esteem scale: {S}elf-evaluation
  of one's social identity}.
\newblock \bibinfo{journal}{{\em Personality and social psychology bulletin\/}}
  \bibinfo{volume}{18}, \bibinfo{number}{3} (\bibinfo{year}{1992}),
  \bibinfo{pages}{302--318}.
\newblock


\bibitem[\protect\citeauthoryear{Modi and Strode}{Modi and Strode}{2020}]%
        {modi2020leadership}
\bibfield{author}{\bibinfo{person}{Sunila Modi} {and} \bibinfo{person}{Diane
  Strode}.} \bibinfo{year}{2020}\natexlab{}.
\newblock \showarticletitle{Leadership in Agile Software Development: {A}
  Systematic Literature Review}. In \bibinfo{booktitle}{{\em ACIS 2020
  Proceedings}}. \bibinfo{publisher}{Springer}, \bibinfo{pages}{55}.
\newblock


\bibitem[\protect\citeauthoryear{Moe, Dings{\o}yr, and Dyb{\aa}}{Moe
  et~al\mbox{.}}{2009a}]%
        {moe2009overcoming}
\bibfield{author}{\bibinfo{person}{N Moe}, \bibinfo{person}{T Dings{\o}yr},
  {and} \bibinfo{person}{T Dyb{\aa}}.} \bibinfo{year}{2009}\natexlab{a}.
\newblock \showarticletitle{Overcoming barriers to self-management in software
  teams}.
\newblock \bibinfo{journal}{{\em IEEE Software\/}} \bibinfo{volume}{26},
  \bibinfo{number}{6} (\bibinfo{year}{2009}), \bibinfo{pages}{20--26}.
\newblock


\bibitem[\protect\citeauthoryear{Moe, Aurum, and Dyb{\aa}}{Moe
  et~al\mbox{.}}{2012}]%
        {moe2012challenges}
\bibfield{author}{\bibinfo{person}{Nils~Brede Moe}, \bibinfo{person}{Ayb{\"u}ke
  Aurum}, {and} \bibinfo{person}{Tore Dyb{\aa}}.}
  \bibinfo{year}{2012}\natexlab{}.
\newblock \showarticletitle{Challenges of shared decision-making: A multiple
  case study of agile software development}.
\newblock \bibinfo{journal}{{\em Information and Software Technology\/}}
  \bibinfo{volume}{54}, \bibinfo{number}{8} (\bibinfo{year}{2012}),
  \bibinfo{pages}{853--865}.
\newblock


\bibitem[\protect\citeauthoryear{Moe, Dings{\o}yr, and Dyb{\aa}}{Moe
  et~al\mbox{.}}{2008}]%
        {moe2008understanding}
\bibfield{author}{\bibinfo{person}{Nils~Brede Moe}, \bibinfo{person}{Torgeir
  Dings{\o}yr}, {and} \bibinfo{person}{Tore Dyb{\aa}}.}
  \bibinfo{year}{2008}\natexlab{}.
\newblock \showarticletitle{Understanding self-organizing teams in agile
  software development}. In \bibinfo{booktitle}{{\em 19th Australian Conference
  on Software Engineering (aswec 2008)}}. \bibinfo{publisher}{IEEE},
  \bibinfo{pages}{76--85}.
\newblock


\bibitem[\protect\citeauthoryear{Moe, Dingsyr, and Kvangardsnes}{Moe
  et~al\mbox{.}}{2009b}]%
        {moe2009understanding}
\bibfield{author}{\bibinfo{person}{Nils~Brede Moe}, \bibinfo{person}{T
  Dingsyr}, {and} \bibinfo{person}{O Kvangardsnes}.}
  \bibinfo{year}{2009}\natexlab{b}.
\newblock \showarticletitle{Understanding shared leadership in agile
  development: A case study}. In \bibinfo{booktitle}{{\em 2009 42nd Hawaii
  International Conference on System Sciences}}. \bibinfo{publisher}{IEEE},
  \bibinfo{pages}{1--10}.
\newblock


\bibitem[\protect\citeauthoryear{Ngwenyama and Nielsen}{Ngwenyama and
  Nielsen}{2003}]%
        {ngwenyama2003competing}
\bibfield{author}{\bibinfo{person}{Ojelanki Ngwenyama} {and}
  \bibinfo{person}{Peter~Axel Nielsen}.} \bibinfo{year}{2003}\natexlab{}.
\newblock \showarticletitle{Competing values in software process improvement:
  an assumption analysis of CMM from an organizational culture perspective}.
\newblock \bibinfo{journal}{{\em IEEE Transactions on Engineering
  Management\/}} \bibinfo{volume}{50}, \bibinfo{number}{1}
  (\bibinfo{year}{2003}), \bibinfo{pages}{100--112}.
\newblock


\bibitem[\protect\citeauthoryear{Przybilla, Pr{\"a}g, Wiesche, and
  Krcmar}{Przybilla et~al\mbox{.}}{2020}]%
        {przybilla2020conceptual}
\bibfield{author}{\bibinfo{person}{Leonard Przybilla},
  \bibinfo{person}{Alexander Pr{\"a}g}, \bibinfo{person}{Manuel Wiesche}, {and}
  \bibinfo{person}{Helmut Krcmar}.} \bibinfo{year}{2020}\natexlab{}.
\newblock \showarticletitle{A Conceptual Model of Antecedents of Emergent
  Leadership in Agile Teams}. In \bibinfo{booktitle}{{\em Proceedings of the
  2020 Computers and People Research Conference}}. \bibinfo{pages}{164--165}.
\newblock


\bibitem[\protect\citeauthoryear{Przybilla, Wiesche, and Krcmar}{Przybilla
  et~al\mbox{.}}{2019}]%
        {przybilla2019emergent}
\bibfield{author}{\bibinfo{person}{Leonard Przybilla}, \bibinfo{person}{Manuel
  Wiesche}, {and} \bibinfo{person}{Helmut Krcmar}.}
  \bibinfo{year}{2019}\natexlab{}.
\newblock \showarticletitle{Emergent Leadership in Agile Teams--an Initial
  Exploration}. In \bibinfo{booktitle}{{\em Proceedings of the 2019 Computers
  and People Research Conference}}. \bibinfo{pages}{176--179}.
\newblock


\bibitem[\protect\citeauthoryear{Ralph}{Ralph}{2018}]%
        {ralph2018re}
\bibfield{author}{\bibinfo{person}{Paul Ralph}.}
  \bibinfo{year}{2018}\natexlab{}.
\newblock \showarticletitle{Re-imagining a course in software project
  management}. In \bibinfo{booktitle}{{\em Proceedings of the 40th
  International Conference on Software Engineering: Software Engineering
  Education and Training}}. \bibinfo{pages}{116--125}.
\newblock


\bibitem[\protect\citeauthoryear{Ralph, Baltes, Bianculli, Dittrich, Felderer,
  Feldt, Filieri, Furia, Graziotin, He, et~al\mbox{.}}{Ralph et~al\mbox{.}}{}]%
        {ralph2020acm}
\bibfield{author}{\bibinfo{person}{Paul Ralph}, \bibinfo{person}{Sebastian
  Baltes}, \bibinfo{person}{Domenico Bianculli}, \bibinfo{person}{Yvonne
  Dittrich}, \bibinfo{person}{Michael Felderer}, \bibinfo{person}{Robert
  Feldt}, \bibinfo{person}{Antonio Filieri}, \bibinfo{person}{Carlo~Alberto
  Furia}, \bibinfo{person}{Daniel Graziotin}, \bibinfo{person}{Pinjia He},
  {and} \bibinfo{person}{others}.}
\newblock \showarticletitle{Empirical Standards for Software Engineering
  Research}.
\newblock  (\bibinfo{year}{????}).
\newblock


\bibitem[\protect\citeauthoryear{Ringstad, Dings{\o}yr, and Moe}{Ringstad
  et~al\mbox{.}}{2011}]%
        {ringstad2011agile}
\bibfield{author}{\bibinfo{person}{Mats~Angermo Ringstad},
  \bibinfo{person}{Torgeir Dings{\o}yr}, {and} \bibinfo{person}{Nils~Brede
  Moe}.} \bibinfo{year}{2011}\natexlab{}.
\newblock \showarticletitle{Agile process improvement: diagnosis and planning
  to improve teamwork}. In \bibinfo{booktitle}{{\em European Conference on
  Software Process Improvement}}. \bibinfo{publisher}{Springer},
  \bibinfo{pages}{167--178}.
\newblock


\bibitem[\protect\citeauthoryear{Robson}{Robson}{2013}]%
        {sean}
\bibfield{author}{\bibinfo{person}{Sean Robson}.}
  \bibinfo{year}{2013}\natexlab{}.
\newblock \bibinfo{booktitle}{{\em Agile {SAP}: {I}ntroducing Flexibility,
  Transparency and Speed to {SAP} Implementations}}.
\newblock \bibinfo{publisher}{IT Governance Publishing},
  \bibinfo{address}{Cambridgeshire}.
\newblock


\bibitem[\protect\citeauthoryear{Schein}{Schein}{2010}]%
        {scheincultlead}
\bibfield{author}{\bibinfo{person}{Edgar Schein}.}
  \bibinfo{year}{2010}\natexlab{}.
\newblock \bibinfo{booktitle}{{\em Organizational culture and leadership\/}
  (\bibinfo{edition}{4} ed.)}.
\newblock \bibinfo{publisher}{Jossey-Bass}, \bibinfo{address}{San Francisco}.
\newblock
\showISBNx{978-0-470-18586-5}


\bibitem[\protect\citeauthoryear{Schwaber}{Schwaber}{2004}]%
        {schwaber}
\bibfield{author}{\bibinfo{person}{Ken Schwaber}.}
  \bibinfo{year}{2004}\natexlab{}.
\newblock \bibinfo{booktitle}{{\em Agile project management with Scrum}}.
\newblock \bibinfo{publisher}{Microsoft Press}, \bibinfo{address}{Redmond,
  Wash.}
\newblock
\showISBNx{0-7356-1993-X}


\bibitem[\protect\citeauthoryear{Schwaber and Sutherland}{Schwaber and
  Sutherland}{2020}]%
        {schwaber2020ScrumGuide}
\bibfield{author}{\bibinfo{person}{Ken Schwaber} {and} \bibinfo{person}{Jeff
  Sutherland}.} \bibinfo{year}{2020}\natexlab{}.
\newblock \bibinfo{title}{The 2020 Scrum Guide}.
\newblock   (\bibinfo{year}{2020}).
\newblock
\showURL{%
\url{https://www.scrumguides.org/scrum-guide.html}}


\bibitem[\protect\citeauthoryear{Sedano, Ralph, and P{\'e}raire}{Sedano
  et~al\mbox{.}}{2020}]%
        {sedano2020dual}
\bibfield{author}{\bibinfo{person}{Todd Sedano}, \bibinfo{person}{Paul Ralph},
  {and} \bibinfo{person}{C{\'e}cile P{\'e}raire}.}
  \bibinfo{year}{2020}\natexlab{}.
\newblock \showarticletitle{Dual-Track Development}.
\newblock \bibinfo{journal}{{\em IEEE Software\/}} \bibinfo{volume}{37},
  \bibinfo{number}{6} (\bibinfo{year}{2020}), \bibinfo{pages}{58--65}.
\newblock


\bibitem[\protect\citeauthoryear{Shih and Huang}{Shih and Huang}{2010}]%
        {shih2010exploring}
\bibfield{author}{\bibinfo{person}{Chiao-Ching Shih} {and}
  \bibinfo{person}{Sun-Jen Huang}.} \bibinfo{year}{2010}\natexlab{}.
\newblock \showarticletitle{Exploring the relationship between organizational
  culture and software process improvement deployment}.
\newblock \bibinfo{journal}{{\em Information \& Management\/}}
  \bibinfo{volume}{47}, \bibinfo{number}{5-6} (\bibinfo{year}{2010}),
  \bibinfo{pages}{271--281}.
\newblock


\bibitem[\protect\citeauthoryear{So and Scholl}{So and Scholl}{2009}]%
        {so}
\bibfield{author}{\bibinfo{person}{Chaehan So} {and} \bibinfo{person}{Wolfgang
  Scholl}.} \bibinfo{year}{2009}\natexlab{}.
\newblock \showarticletitle{Perceptive agile measurement: {N}ew instruments for
  quantitative studies in the pursuit of the social-psychological effect of
  agile practices}.
\newblock In \bibinfo{booktitle}{{\em Agile Processes in Software Engineering
  and Extreme Programming}}. \bibinfo{publisher}{Springer},
  \bibinfo{pages}{83--93}.
\newblock


\bibitem[\protect\citeauthoryear{Spiegler, Graziotin, Heinecke, and
  Wagner}{Spiegler et~al\mbox{.}}{2020}]%
        {spiegler2020quantitative}
\bibfield{author}{\bibinfo{person}{Simone~V Spiegler}, \bibinfo{person}{Daniel
  Graziotin}, \bibinfo{person}{Christoph Heinecke}, {and}
  \bibinfo{person}{Stefan Wagner}.} \bibinfo{year}{2020}\natexlab{}.
\newblock \showarticletitle{A Quantitative Exploration of the 9-Factor Theory:
  {D}istribution of Leadership Roles Between Scrum Master and Agile Team}. In
  \bibinfo{booktitle}{{\em International Conference on Agile Software
  Development}}. \bibinfo{publisher}{Springer}, \bibinfo{pages}{162--177}.
\newblock


\bibitem[\protect\citeauthoryear{Spiegler, Heinecke, and Wagner}{Spiegler
  et~al\mbox{.}}{2021}]%
        {spiegler2021empirical}
\bibfield{author}{\bibinfo{person}{Simone~V Spiegler},
  \bibinfo{person}{Christoph Heinecke}, {and} \bibinfo{person}{Stefan Wagner}.}
  \bibinfo{year}{2021}\natexlab{}.
\newblock \showarticletitle{An empirical study on changing leadership in agile
  teams}.
\newblock \bibinfo{journal}{{\em Empirical Software Engineering\/}}
  \bibinfo{volume}{26}, \bibinfo{number}{3} (\bibinfo{year}{2021}),
  \bibinfo{pages}{1--35}.
\newblock


\bibitem[\protect\citeauthoryear{Srivastava and Jain}{Srivastava and
  Jain}{2017}]%
        {srivastava2017leadership}
\bibfield{author}{\bibinfo{person}{Pallavi Srivastava} {and}
  \bibinfo{person}{Shilpi Jain}.} \bibinfo{year}{2017}\natexlab{}.
\newblock \showarticletitle{A leadership framework for distributed
  self-organized scrum teams}.
\newblock \bibinfo{journal}{{\em Team Performance Management\/}}
  \bibinfo{volume}{23}, \bibinfo{number}{5/6} (\bibinfo{year}{2017}),
  \bibinfo{pages}{293--314}.
\newblock


\bibitem[\protect\citeauthoryear{Strode, Huff, and Tretiakov}{Strode
  et~al\mbox{.}}{2009}]%
        {strode2009impact}
\bibfield{author}{\bibinfo{person}{Diane~E Strode}, \bibinfo{person}{Sid~L
  Huff}, {and} \bibinfo{person}{Alexei Tretiakov}.}
  \bibinfo{year}{2009}\natexlab{}.
\newblock \showarticletitle{The impact of organizational culture on agile
  method use}. In \bibinfo{booktitle}{{\em 2009 42nd Hawaii International
  Conference on System Sciences}}. \bibinfo{publisher}{IEEE},
  \bibinfo{pages}{1--9}.
\newblock


\bibitem[\protect\citeauthoryear{Teh, Baniassad, Van~Rooy, and Boughton}{Teh
  et~al\mbox{.}}{2012}]%
        {teh}
\bibfield{author}{\bibinfo{person}{Alvin Teh}, \bibinfo{person}{Elisa
  Baniassad}, \bibinfo{person}{Dirk Van~Rooy}, {and} \bibinfo{person}{Clive
  Boughton}.} \bibinfo{year}{2012}\natexlab{}.
\newblock \showarticletitle{Social Psychology and Software Teams:
  {E}stablishing Task-Effective Group Norms}.
\newblock \bibinfo{journal}{{\em IEEE Software\/}} \bibinfo{volume}{29},
  \bibinfo{number}{4} (\bibinfo{year}{2012}), \bibinfo{pages}{53--58}.
\newblock


\bibitem[\protect\citeauthoryear{Tolfo and Wazlawick}{Tolfo and
  Wazlawick}{2008}]%
        {tolfo2008}
\bibfield{author}{\bibinfo{person}{Cristiano Tolfo} {and} \bibinfo{person}{Raul
  Wazlawick}.} \bibinfo{year}{2008}\natexlab{}.
\newblock \showarticletitle{The influence of organizational culture on the
  adoption of extreme programming}.
\newblock \bibinfo{journal}{{\em Journal of systems and software\/}}
  \bibinfo{volume}{81}, \bibinfo{number}{11} (\bibinfo{year}{2008}),
  \bibinfo{pages}{1955--1967}.
\newblock


\bibitem[\protect\citeauthoryear{Tolstoy}{Tolstoy}{1869}]%
        {tolstoy}
\bibfield{author}{\bibinfo{person}{L. Tolstoy}.}
  \bibinfo{year}{1869}\natexlab{}.
\newblock \bibinfo{booktitle}{{\em War and Peace}}.
\newblock \bibinfo{publisher}{Penguin}, \bibinfo{address}{Harmondsworth, UK}.
\newblock


\bibitem[\protect\citeauthoryear{Tuckman and Jensen}{Tuckman and
  Jensen}{1977}]%
        {tuckman}
\bibfield{author}{\bibinfo{person}{Bruce Tuckman} {and} \bibinfo{person}{Mary
  Jensen}.} \bibinfo{year}{1977}\natexlab{}.
\newblock \showarticletitle{Stages of small-group development revisited}.
\newblock \bibinfo{journal}{{\em Group \& Organization Management\/}}
  \bibinfo{volume}{2}, \bibinfo{number}{4} (\bibinfo{year}{1977}),
  \bibinfo{pages}{419--427}.
\newblock


\bibitem[\protect\citeauthoryear{Veiseh, Eghbali, et~al\mbox{.}}{Veiseh
  et~al\mbox{.}}{2014}]%
        {veiseh2014study}
\bibfield{author}{\bibinfo{person}{Seidmehdi Veiseh}, \bibinfo{person}{Neeman
  Eghbali}, {and} \bibinfo{person}{others}.} \bibinfo{year}{2014}\natexlab{}.
\newblock \showarticletitle{A study on ranking the effects of transformational
  leadership style on organizational agility and mediating role of
  organizational creativity}.
\newblock \bibinfo{journal}{{\em Management Science Letters\/}}
  \bibinfo{volume}{4}, \bibinfo{number}{9} (\bibinfo{year}{2014}),
  \bibinfo{pages}{2121--2128}.
\newblock


\bibitem[\protect\citeauthoryear{Von~Cranach}{Von~Cranach}{1986}]%
        {von1986leadership}
\bibfield{author}{\bibinfo{person}{Mario Von~Cranach}.}
  \bibinfo{year}{1986}\natexlab{}.
\newblock \showarticletitle{Leadership as a function of group action}.
\newblock In \bibinfo{booktitle}{{\em Changing conceptions of leadership}},
  \bibfield{editor}{\bibinfo{person}{Carl~F. Graumann} {and}
  \bibinfo{person}{Serge Moscovici}} (Eds.). \bibinfo{publisher}{Springer},
  \bibinfo{address}{New York, US}, \bibinfo{pages}{115--134}.
\newblock


\bibitem[\protect\citeauthoryear{Wang, Oh, Courtright, and Colbert}{Wang
  et~al\mbox{.}}{2011}]%
        {wang2011transformational}
\bibfield{author}{\bibinfo{person}{Gang Wang}, \bibinfo{person}{In-Sue Oh},
  \bibinfo{person}{Stephen~H Courtright}, {and} \bibinfo{person}{Amy~E
  Colbert}.} \bibinfo{year}{2011}\natexlab{}.
\newblock \showarticletitle{Transformational leadership and performance across
  criteria and levels: A meta-analytic review of 25 years of research}.
\newblock \bibinfo{journal}{{\em Group \& organization management\/}}
  \bibinfo{volume}{36}, \bibinfo{number}{2} (\bibinfo{year}{2011}),
  \bibinfo{pages}{223--270}.
\newblock


\bibitem[\protect\citeauthoryear{Wheelan and Hochberger}{Wheelan and
  Hochberger}{1996}]%
        {wheelan}
\bibfield{author}{\bibinfo{person}{S. Wheelan} {and} \bibinfo{person}{J.
  Hochberger}.} \bibinfo{year}{1996}\natexlab{}.
\newblock \showarticletitle{Validation studies of the group development
  questionnaire}.
\newblock \bibinfo{journal}{{\em Small Group Research\/}} \bibinfo{volume}{27},
  \bibinfo{number}{1} (\bibinfo{year}{1996}), \bibinfo{pages}{143--170}.
\newblock


\bibitem[\protect\citeauthoryear{Williams}{Williams}{2012}]%
        {williams}
\bibfield{author}{\bibinfo{person}{Laurie Williams}.}
  \bibinfo{year}{2012}\natexlab{}.
\newblock \showarticletitle{What agile teams think of agile principles}.
\newblock \bibinfo{journal}{{\it Commun. ACM}} \bibinfo{volume}{55},
  \bibinfo{number}{4} (\bibinfo{year}{2012}), \bibinfo{pages}{71--76}.
\newblock


\bibitem[\protect\citeauthoryear{Wufka and Ralph}{Wufka and Ralph}{2015}]%
        {wufka2015explaining}
\bibfield{author}{\bibinfo{person}{Michael Wufka} {and} \bibinfo{person}{Paul
  Ralph}.} \bibinfo{year}{2015}\natexlab{}.
\newblock \showarticletitle{Explaining Agility with a Process Theory of
  Change}. In \bibinfo{booktitle}{{\em Agile Conference (AGILE)}}.
  \bibinfo{publisher}{IEEE}, \bibinfo{pages}{60--64}.
\newblock


\end{thebibliography}

\end{document}